\documentclass[twocolumn,prc,amssymb, aps,superscriptaddress, 
amsmath,floatfix]{revtex4}

\usepackage{color}
\usepackage{amsmath,mathtools,booktabs,
            microtype} 

\usepackage{epstopdf}
\usepackage{graphics}
\usepackage{graphicx}
\usepackage{dcolumn}
\usepackage{bm}
\usepackage{epsfig}
\usepackage{epstopdf}
\usepackage{times}
\usepackage{url}
\usepackage[version=4]{mhchem}

\newcommand{\be}{\begin{equation}}
\newcommand{\ee}{  \end{equation}}
\newcommand{\ba}{\begin{eqnarray}}
\newcommand{\ea}{  \end{eqnarray}}
\newcommand{\ve}{\varepsilon}

\begin{document}

\title{Laser-nucleus interactions in the sudden regime}

\author{Sergei \surname{Kobzak}}
\email{kobzak@mpi-hd.mpg.de}
\affiliation{Max-Planck-Institut f\"ur Kernphysik, Saupfercheckweg 1, D-69117 Heidelberg, Germany}

\author{Hans A. \surname{Weidenm\"uller}}
\email{haw@mpi-hd.mpg.de}
\affiliation{Max-Planck-Institut f\"ur Kernphysik, Saupfercheckweg 1, D-69117 Heidelberg, Germany}

\author{Adriana \surname{P\'alffy}}
\email{adriana.palffy-buss@fau.de}
\affiliation{Max-Planck-Institut f\"ur Kernphysik, Saupfercheckweg 1, D-69117 Heidelberg, Germany}
\affiliation{Department of Physics, Friedrich-Alexander-Universit\"at  Erlangen-N\"urnberg,  D-91058 Erlangen, Germany}

\date{\today}

\begin{abstract}

The interaction between medium-weight nuclei and a strong zeptosecond
laser pulse of MeV photons is investigated theoretically. Multiple
absorption of photons competes with nuclear equilibration. We
investigate the sudden regime. Here the rate of photon absorption is
so strong that there is no time for the nucleus to fully equilibrate
after each photon absorption process. We follow the temporal evolution
of the system in terms of a set of rate equations. These account for
dipole absorption and induced dipole emission, equilibration (modeled
in terms of particle-hole states coupled by the residual nuclear
interaction), and neutron decay (populating a chain of proton-rich
nuclei). Our results are compared with earlier work addressing the
adiabatic regime where equilibration is instantaneous. We predict the
degree of excitation and the range of nuclei reached by neutron
evaporation.  These findings are relevant for planning future
experiments.

\end{abstract}

\maketitle

\section{Introduction }

Exciting experimental developments at petawatt laser facilities
\cite{danson2019} combined with experimental, computational and
theoretical advances in the production of high-energy laser
pulses~\cite{Esi2009,Kie2009,Mey2009,Mou11,
  Kie13,Bul2013,Mu2013,Li2014} give rise to the hope that intense
pulses with photon energy $\hbar \omega_0$ in the few MeV range and
with a typical energy spread $\sigma$ in the $10$ keV range will
become available in the near future. Efforts in that direction are
presently undertaken at the Nuclear Pillar of the Extreme Light
Infrastructure under construction in Romania~\cite{ELI-web}, and in
the development of Gamma Factories at the Large Hadron Collider of
CERN~\cite{gamma-fact}. How would such a pulse interact with a
nucleus?

For a photon with an energy in the MeV range, the product of photon
wave number $k$ and nuclear radius $R$ obeys $k R \ll 1$. Therefore,
we consider only dipole processes (even though higher multipolarities
might be important for some nuclei at small excitation
energies~\cite{Pal08}). Single dipole absorption excites the nuclear
giant dipole resonance (GDR). In a shell-model picture the GDR is a
superposition of particle-hole excitations out of the ground state and
is not an eigenstate of the nuclear Hamiltonian
$\hat{H}$. These particle-hole excitations actually do
  not all have the same energy. That leads to a spreading of the GDR
  often referred to as Landau damping, a one-body effect \cite{Fiolhais1986,Speth1991}. The residual
  two-body interaction mixes the particle-hole excitations with each
  other and with other shell-model configurations and, thus, spreads
  the GDR over the eigenstates of $\hat{H}$, leading to a Lorentzian
  distribution of the dipole strength with width
  $\Gamma^\downarrow$. For low-lying modes with excitation energies
of up to 10 or 20 MeV the ``spreading width'' $\Gamma^\downarrow$
of the nuclear GDR~\cite{Feshbach1964, Herman1992} has values around
$5$ MeV~\cite{Wei09}. In a time-dependent picture the spreading of the
GDR over the eigenstates of $\hat{H}$ can be viewed as statistical
equilibration~\cite{Agassi1975} with characteristic time scale
$\tau_{\rm eq} = \hbar / \Gamma^\downarrow$. A similar
  order of magnitude for the generic time required to reach thermal
  equilibrium $\tau_{\rm eq} \simeq 10^{-22}$~s can be achieved by
  considering the traversal time for medium-weight nuclei
  \cite{Bortignon1991}.  We note however that the definition of an
  equilibration time becomes more complex once very high excitation
  energies are achieved, for instance in hot GDRs
  \cite{Santonocito2020}, accompanied by strong neutron evaporation
  rates and corresponding short lifetimes \cite{Bortignon1991}.

The strength of dipole absorption is measured by the rate $R_{\rm dip}$ (or, equivalently, by the effective dipole width
$\widetilde{\Gamma}_{\rm dip} = \hbar R_{\rm dip}$ or the time scale
$\tau_{\rm dip} = \hbar / \widetilde{\Gamma}_{\rm dip}$ for dipole
absorption). The standard nuclear dipole width has values in the keV
range. However, for a laser pulse containing $N \gg 1$ photons within
few tens of zs (1~zs$=10^{-21}$~s), that width is boosted by the factor $N$ even when the
pulse is not coherent~\cite{palffy2020}, and the effective dipole
width can easily take values in the MeV range. That makes multiple
dipole absorption of photons out of the same laser pulse a likely
process.

For the following qualitative comparison of
  $\widetilde{\Gamma}_{\rm dip}$ and $ \Gamma^\downarrow$ we consider
  both quantities as independent of excitation energy. That picture is
  only an approximation. Experimental evidence for hot GDR quenching
  \cite{Santonocito2020} was interpreted as an increase of $
  \Gamma^\downarrow$ with temperature, reaching up to 20~MeV (50~MeV)
  at 160~MeV (220~MeV) excitation energy,
  respectively~\cite{Yoshida1990}. Such an increase of $
  \Gamma^\downarrow$ with temperature would lead to a weak decrease of
  the effective dipole absorption width $\widetilde{\Gamma}_{\rm dip}$
  as indicated by the expression for $R_{\rm dip}$
  \cite{palffy2020}. However, for the mere purpose of defining
  laser-nucleus interaction regimes, it is sufficient to compare the
  two widths at small excitation energies, where they can reach
  comparable values depending on the laser gamma-ray parameters.
Once $\widetilde{\Gamma}_{\rm dip} \approx \Gamma^\downarrow$, we
expect that multiple dipole absorption leads to multiple GDR-type
excitations, each accompanied by internal nuclear equilibration. The
ratio of the two competing widths $\Gamma^\downarrow$ and
$\widetilde{\Gamma}_{\rm dip}$ then defines three regimes: (i) the
perturbative regime $\widetilde{\Gamma}_{\rm dip} \ll
\Gamma^\downarrow$, (ii) the quasiadiabatic regime
$\widetilde{\Gamma}_{\rm dip} \leq \Gamma^\downarrow$ and (iii) the
sudden regime $\widetilde{\Gamma}_{\rm dip} \gg \Gamma^\downarrow$.
The perturbative regime (i) was studied in
Refs.~\cite{Die10,Wei11}. The term ``quasiadiabatic'' in (ii) refers
to the assumption that after each photon absorption process, the
nucleus reaches equilibrium prior to the absorption of the next
photon. Theoretical and numerical studies~\cite{Pal14,Pal15} in that
regime are based on a statistical approach and make use of rate
equations. These have shown that multiple photon absorption produces
compound nuclei in the so-far unexplored regime of several hundred MeV
excitation energy and low angular momentum. The nuclei so produced
undergo sequential neutron decay with intermittent further dipole
absorption and equilibration, leading to a chain of highly excited
proton-rich nuclei.

In this paper we address the sudden regime (iii). To model a
situation where after each photoabsorption process there is not
sufficient time for equilibration, we need a detailed description of
the states of the compound nucleus. We use the shell model, assuming
that the ground state of the target nucleus has a doubly-closed
shell. The last occupied single-particle state defines the Fermi
surface. Excited states are multiple particle-hole excitations out of
the ground state (referred to as $m$p-$m$h states with integer $m$).

For that picture, a manageable theoretical framework cannot be
established without statistical assumptions. Particle-hole states are
grouped into classes defined by particle-hole number $m$ and total
energy,  as discussed in more detail in Sec.~\ref{basi}.
  It is assumed that within each class of $m$p-$m$h states, the
  residual interaction is so strong that equilibration is much faster
  than the equilibration between different classes, and can be
  considered to be quasi instantaneous. This assumption is also used
  in precompound reaction models and has proven its validity by good
  agreement with experimental data \cite{Blann1985}. A second related
  assumption is that because of the strong mixing within one class, the
  eigenfunctions of the time-independent Hamiltonian are
  Gaussian-distributed random variables, and the eigenvalues obey
  Wigner-Dyson statistics \cite{Haw2021}. This assumption was
  thoroughly tested in Ref.~\cite{Zelevinsky1996}. These two
  assumptions guarantee that both the matrix elements of the residual
interaction connecting states in different classes and those of the
dipole operator, are zero-centered Gaussian-distributed random
variables~\cite{Agassi1975}. Rates are obtained as mean values over
these distributions. The rates for nuclear equilibration are
proportional to mean values of squares of matrix elements of the
residual interaction connecting states in different classes, and to
the level density of the $m$p-$m$h states reached. The rate for dipole
absorption is similarly proportional to the mean square matrix element
for dipole absorption~\cite{palffy2020} and to the density of final
states. Multiple dipole absorption leads to nuclear excitation far
above yrast. Calculation of the rates requires, therefore, the
knowledge of $m$p-$m$h level densities at high excitation energy (up
to several $100$ MeV) and for large particle numbers. A reliable
approximation for these densities in terms of the single-particle
level density of the shell model was worked out in
Refs.~\cite{Pal13a, Pal13b} and is used in what follows.

The rates are used in rate equations. These describe the time
evolution of the average occupation probabilities of classes of
$m$p-$m$h states under the influence of the external field of the
laser. They account for the following competing processes:
photoabsorption and its inverse process stimulated photon emission,
equilibration, and neutron evaporation. In a manner similar to the
theory of precompound reactions~\cite{Wei08}, equilibration is taken
into account by coupling different $m$p-$m$h classes at the same
energy. Absorption of a photon by an $m$p-$m$h state either generates
an additional particle-hole pair promoting the nucleus to class
$(m+1)$p-$(m+1)$h, or it increases the energy of an existing
particle-hole pair. Conversely, stimulated emission leads to the
annihilation of a particle-hole pair, or it reduces the energy of an
existing particle-hole pair without changing $m$.  We
  disregard here possible collective excitations which could play a
  role at small excitation energies. Neutron evaporation changes mass
number from even to odd and conversely. For odd-mass nuclei we
interpolate between the neighboring even-mass nuclei. We consider,
thus, only states with equal particle-hole numbers. We neglect
particle loss from direct photon excitation of particles (protons or
neutrons) into continuum states. Thus we confine ourselves to a chain
of nuclei with equal proton numbers. Ensuing limitations and possible
corrections have been addressed qualitatively in Ref.~\cite{Pal15} for
the quasiadiabatic regime. The relevance of these processes for the
deep sudden regime is briefly addressed in the concluding remarks of
this paper in Sec.~\ref{disc}. We simplify the treatment of the
problem by disregarding spin altogether. That was justified in
Ref.~\cite{Pal15} by the slow increase of total spin value with
multiple photon absorption.

We consider the interaction of a strong zeptosecond laser pulse with a medium-weight nucleus with mass
number $A$. For $\widetilde{\Gamma}_{\rm dip}$ we use values in the
range $1$ -- $20$ MeV. In the course of the reaction, up to $N_0 \approx
140$ photons may be absorbed. We neglect the resulting reduction of $N
\to N - N_0$ in the boost factor of $\widetilde{\Gamma}_{\rm
  dip}$. The energy $\hbar \omega_0$ per photon is $5$ MeV, and the
duration of the pulse is $\tau=\hbar / \sigma$ where $\sigma$ is of
the order of several $10$ keV so that $\tau\approx 10^{- 20}$~s. We
investigate the temporal evolution of the nucleus over the laser pulse duration, and we follow the chain of neutron
evaporation processes towards proton-rich nuclei. Fission is expected
to be important only for very heavy nuclei and is disregarded. To
illustrate the role of the equilibration process, we compare results
for the sudden and for the quasiadiabatic regime. In the absence of
nucleon emission and fission, photon absorption would saturate at an
excitation energy where the rates for absorption and for stimulated
emission become equal. That energy is given by the maximum of the
total level density summed over all particle-hole classes. The larger
the effective dipole absorption rate, the faster this saturation
energy is reached. Neutron evaporation takes over at an energy below
the saturation point. The combination of repeated neutron emission and
continued dipole absorption by the daughter nuclei then produces
proton-rich nuclei far from the valley of stability. This picture is
qualitatively similar to but quantitatively somewhat different from the
results for the quasiadiabatic regime.

The paper is structured as follows. The rate equation and the
transition rates are introduced in Sec.~\ref{mas}.  This section also
addresses the densities of accessible states for $m$p-$m$h classes.
Numerical results follow in Sec.~\ref{numres} and the paper concludes
with a discussion in Sec.~\ref{disc}.

\section{Rate Equations}
\label{mas}

\subsection{Basic Approach}
\label{basi}

With $A$ the even mass number of the target nucleus, we consider a
chain of $(n + 1)$ nuclei with mass numbers $A - i$ where $i = 0, 1,
2, \ldots, n$, with an arbitrary cutoff at $i = n$. In the target
($i=0$),  absorption of $k$ laser photons will increase
  the excitation energy by $k\hbar\omega_0$ and potentially also
  change the particle-hole number.  We group the nuclear states
  according to the generation $i$, the particle-hole number $m$, and
  the total energy, which for our case will be a multiple of the laser
  photon energy $\hbar\omega_0$. In the following we therefore use
  classes labeled $(i, k, m)$. The equilibration processes between
  classes are discussed below. The level density in each class is
$\rho_m(0, k)$. Single or multiple neutron decay of the target
populates an energy continuum of states in the daughter nuclei labeled
$i = 1, 2, \ldots, n$. For even-mass daughter nuclei $i$, the
$m$p-$m$h states in the energy interval between $(k - 1/2) \hbar
\omega_0$ and $(k + 1/2) \hbar \omega_0$ form class $(i, k, m)$. The
class of particle-hole states with excitation energies in the interval
$0 \leq E \leq (1/2) \hbar \omega_0$ is labeled $(i, 0, m)$. The
average level density of the states in class $(i, k, m)$ is denoted by
$\rho_m(i, k)$. For odd-mass daughter nuclei we use energy intervals
$k$ defined in the same manner. We avoid introducing $m$p-$(m \pm 1)$h
states and their level densities and use a simplification instead. We
neglect the even-odd staggering of the ground-state energies as well
as that of the spin-cutoff factor, and we approximate the level
density for odd $i$ by interpolating between the values for the two
neighboring even-mass nuclei. In other words, we use the expression
for the level density for even mass numbers given in
Ref.~\cite{Pal13b} indiscriminately for both even and odd $A$.

The rate equation for the average total occupation probability $P_m(i,
k, t)$ of the states in class $(i, k, m)$ as a function of time $t$ is
\begin{widetext}
\begin{eqnarray}
\label{eq:me3}
\dot{P}_m(i, k, t) &=& \sum_{m'=m \pm  1} V^2_{m' m}(i, k) \rho_m(i, k)
P_{m'}(i, k, t) - \sum_{m'=m \pm  1} V^2_{m m'}(i, k) \rho_{m'}(i, k)
P_{m}(i, k, t)  \nonumber \\
&+& \Theta(\tau - t)  \sum_{m'=m,m\pm 1} \bigg\{  W^2_{k-1k;m' m}(i)
\rho_m(i, k) P_{m'}(i, k - 1, t) + W^2_{k+1k;m' m}(i) \rho_m(i, k)
P_{m'}(i, k + 1, t)  \nonumber \\
&- &W^2_{kk+1;m m'}(i) \rho_{m'}(i, k + 1) P_{m}(i, k, t) 
- W^2_{kk-1;m m'} (i)\rho_{m'}(i, k - 1) P_{m}(i, k, t) \bigg\}
\nonumber \\
&-& \Gamma_{\rm N}(i, k, m) P_m(i, k, t) + \sum_{\substack{k'\\ m'= m, m + 1}}
\Gamma_{\rm N}(i-1, k'\to k, m' \to m)P_{m'}(i-1, k', t)  \ .
\end{eqnarray}
\end{widetext}
We have put $\hbar = 1$. The dot denotes the time derivative. The
equation takes into account three processes: {\bf (i)} equilibration
of occupation probability of the different $m$p-$m$h classes at
constant energy (first line); {\bf (ii)} dipole excitation and
stimulated dipole emission by the MeV laser pulse (second and third
line); and {\bf (iii)} neutron evaporation populating nucleus $A - i -
1$ at the expense of nucleus $A - i$ (last line, where we have defined
$P_m(- 1, k, t) = 0$). The Heaviside function $\Theta$ accounts for
the fact that process {\bf (ii)} occurs only for the duration time $\tau$ of the laser pulse. The initial condition is $P_m(i, k, 0) =
\delta_{i 0} \delta_{k 0}\delta_{m0}$.

\begin{figure}
\centering
\includegraphics[width=\linewidth]{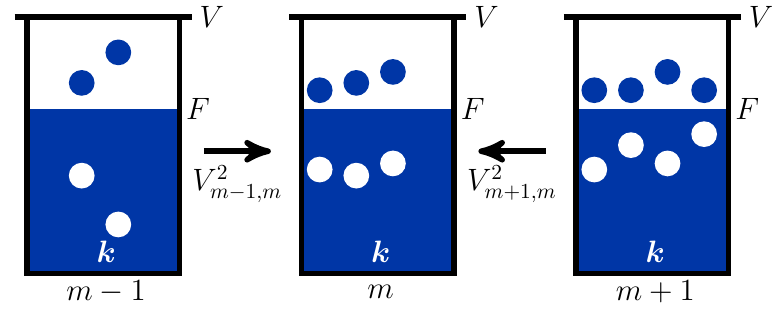}
\caption{Transitions into class \(m\) from neighboring classes ${m\pm 1}$ described by the nucleon-nucleon interaction matrix element $V_{m' m}(i, k)$. Here $F$ represents the Fermi energy and $V$ the threshold
  energy of the single-particle potential. Particles (blue filled
  circles) are above $F$; holes (white full circles) are below $F$.}
\label{v}
\end{figure}

In each nucleus $i$, the equilibration process {\bf (i)} involves the
coupling of classes $(i, k, m)$ at fixed energy $k \hbar \omega_0$ by
the residual interaction. The rate is given by $V_{m' m}^2 \rho_m(i,
k)$, with $V^2_{m m'} = V^2_{m' m}$ the mean square matrix element.
 We recall here our basic picture: Dipole absorption
  primarily populates distinct particle-hole states with somewhat
  different energies (Landau damping) \cite{Fiolhais1986,Speth1991}. We assume that the states
  within the same $m$p-$m$h class are quickly mixed by the residual
  two-body interaction. The remaining part of the residual two-body
  interaction mixes classes of states with different $m$ with rate
  $V_{m' m}^2 \rho_m(i, k)$. Obviously only neighboring classes $m'= m
  \pm 1$ (see Fig.~\ref{v}) are coupled. The inverse of the total time
  needed for such mixing equals $\Gamma^\downarrow / \hbar$. That
  picture is supported by the temperature dependence of the hot GDR 
  width  which is interpreted as
  being from two-body collisions \cite{Smerzi1991,Santonocito2020}. Class $(i, k, m)$ may gain (lose)
occupation probability because of feeding from (depletion to) classes $(i,
k, m \pm 1)$, respectively. Equilibrium is reached when $P_{m}(i, k,
t) \propto c \rho_{m}(i, k)$ with a constant $c$ independent of $m$.

\begin{figure}
\centering
\includegraphics[width=\linewidth]{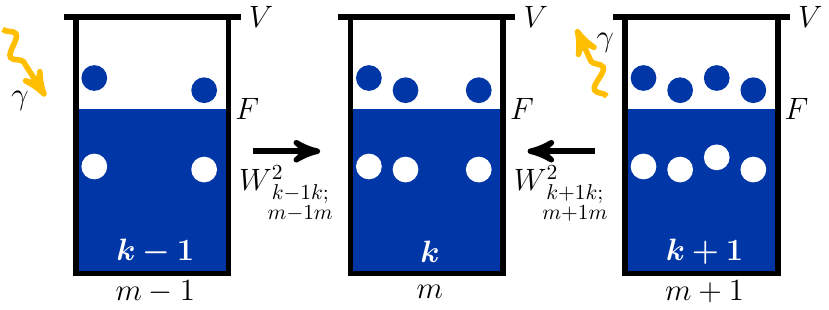}
\caption{Transitions into class $(i, k, m)$ from neighboring classes
  $(i, k - 1, m - 1)$ and $(i, k + 1, m + 1)$ owing to the laser-nucleus
  interaction matrix element $W_{k' k; m' m}(i)$ with $m'= m \pm
  1$. See Fig.~\ref{v} for further notation.}
\label{w}
\end{figure}

\begin{figure}
\centering
\includegraphics[width=\linewidth]{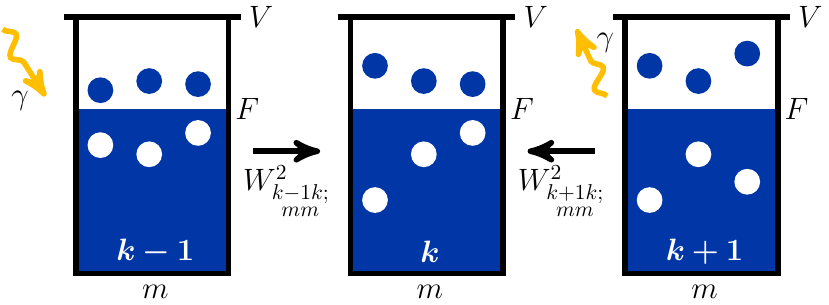}
\caption{ Transitions into class $(i, k, m)$ from neighboring classes
  $(i, k - 1, m)$ and $(i, k + 1, m)$ owing to the laser-nucleus
  interaction matrix element $W_{k' k;m m}(i)$. See Fig.~\ref{v} for
  further notation.}
\label{ww}
\end{figure}

For processes {\bf (ii)}, the class $(i, k, m)$ is fed by coherent dipole
excitation of classes $(i, k - 1, m)$ and $(i, k - 1, m - 1)$ and by
stimulated dipole emission from classes $(i, k + 1, m + 1)$ and $(i, k
+ 1, m)$. Class $(i, k, m)$ is depleted by dipole absorption exciting
classes $(i, k + 1, m)$ and $(i, k + 1, m + 1)$, and by stimulated
dipole emission to classes $(i, k - 1, m)$ and $(i, k - 1, m - 1)$.
Processes where dipole transitions change (do not change)
particle-hole number are illustrated in Fig.~\ref{w} (in
Fig.~\ref{ww}, respectively). The rates feeding class $(i, k, m)$ are
written as $W_{k' k; m' m}^2(i) \rho_m(i, k)$ with $k' = k - 1,\, m' =
m, m - 1$ and $k' = k + 1, \, m' = m, m + 1$. Here $W^2_{k k';m m'}(i) =
W^2_{k' k;m'm}(i)$ is the average square of the transition matrix
element. We have simplified the notation by summing $m'$
indiscriminately over $m$ and $m \pm 1$. That requires that we set
$W_{k - 1 k; m + 1 m}^2(i) = 0 = W_{k + 1 k; m - 1 m}^2(i)$.

The neutron decay process {\bf (iii)} depletes the states in class
$(i, k, m)$ at the rate $\Gamma_N(i, k, m)$. Neutron decay of the
states $(i - 1, k', m')$ in the parent nucleus with mass number $A + 1
- i$ feeds the states in class $(i, k, m)$ with the rate $\Gamma_N(i -
1, k' \to k,m'\to m)$. We allow only for $m = m', m' - 1$.

The set of rate equations~(\ref{eq:me3}) is similar in spirit to but
much more involved than the master equation solved in the
quasiadiabatic case~\cite{Pal14,Pal15}. There, equilibration was
assumed from the outset. At fixed excitation energy only the total
occupation probability and the total level density (both summed over
all $m$) come into play. There are no particle-hole classes. In our
case, separate treatment of the particle-hole classes obviously
increases the number of coupled differential equations significantly.

\subsection{Transition Rates}
\label{rates}

In this section we give expressions for the rates of the three
processes. We mention in passing that for the adiabatic case the rates
for processes {\bf (ii)} and {\bf (iii)} have been defined,
calculated, and discussed in Refs.~\cite{Pal14,Pal15}.  Because $\hbar =
1$ we use the expressions ``width'' and ``rate'' interchangeably.

\subsubsection{Equilibration rate \label{eq-sec}}

Equilibration is the result of the coupling of neighboring
particle-hole classes at constant energy. To estimate that coupling we
use the optical model (see Ref.~\cite{Herman1992}). The imaginary part
$W(\varepsilon)$ of the optical model potential for nucleons accounts for
two-body collisions that remove a nucleon at energy $\varepsilon$
above the Fermi energy from the incident channel and create a
$2$p-$1$h state. As function of time, the occupation probability in
the incident channel decreases exponentially as $\exp \{ - 2
W(\varepsilon) t \}$. Therefore, we identify $2 W(\varepsilon)$ with
the spreading width of a (quasiparticle) nucleon above the Fermi
surface. The concept of the optical model applies also to hole states,
with $\varepsilon$ now the energy of the hole, i.e., the energy below
the Fermi energy. Each particle and each hole in an $m$p-$m$h state at
energy $E$ may undergo a two-body collision leading to an $(m +
1)$p-$(m + 1)$h state. The total spreading width for such a particle
(hole) is obtained by averaging the optical model over the normalized
probability $D_{\mathrm{p}}(m, \varepsilon, E)$ for finding the
particle (hole) at energy $\varepsilon$ in the $m$p-$m$h state at
energy $E$. The total spreading width for all $m$ particles (holes) is
obtained by multiplying the result by $m$. Thus,
\begin{eqnarray}
\label{eq:gamma_spread}
\Gamma_{m \to m+1}^{\downarrow}&=& 2m \int \limits _{0}^{V-F}
D_{\mathrm{p}}(m,\varepsilon, E) W(\varepsilon) d \varepsilon \nonumber \\
&+& 2 m \int \limits_{0}^{F} D_{\mathrm{h}}(m,\varepsilon, E) W(\varepsilon)
d \varepsilon \ .
\end{eqnarray}
We recall that $V$ is the threshold energy of the shell-model
potential and $F$ the Fermi energy, respectively. The distributions
are given by~\cite{Herman1992}
\begin{equation}
\begin{array}{l}
  {D_{\mathrm{p}}(m,\varepsilon, E)=K_{\mathrm{p}} \dfrac{\tilde\rho_{m-1, m}
      (i, E-\varepsilon)}{\rho_{m}(i, E)}} \ , \\
  {D_{\mathrm{h}}(m,\varepsilon, E)=K_{\mathrm{h}} \dfrac{\tilde\rho_{m,m-1}
      (i, E-\varepsilon)}{\rho_{m}(i, E)}} \ .
\end{array}
\end{equation}
Here \(K_\mathrm{p}\) and \(K_\mathrm{h}\) are normalization
constants, and $\tilde\rho_{p,h}(i, E)$ with $p \neq h$ is the density
of $p$p-$h$h states at energy $E$. Finally we use
\begin{equation}
  \Gamma_{m \to m + 1}^{\downarrow} = 2 \pi {V_{m m + 1}^{2}}(i,k) \rho_{m + 1}
  (i,k) \ .
\end{equation}
For the process $m \to (m - 1)$ we use \(V^2_{m m'} = V^2_{m' m}\) and
detailed balance so that
\begin{equation}
  \label{end}
  \Gamma_{m \to m - 1}^{\downarrow} = \Gamma_{m - 1 \to m}^{\downarrow}
  \frac{\rho_{m-1}(i,k)}{\rho_{m}(i,k)} \ .
\end{equation}
 Following Refs.~\cite{Mahaux1985,Herman1992}, we use \(W(\varepsilon)
= c \varepsilon ^2\), with $c=0.003$~MeV$^{-1}$. Further employing
Eqs.~(\ref{eq:gamma_spread}) -- (\ref{end}), and the level densities
given in Sec.~\ref{dens} below, we arrive at the numerical values
for the rates used in Eqs.~(\ref{eq:me3}).

\subsubsection{Dipole transitions}

The effective dipole width for excitation starting from the ground
state is given by $\widetilde{\Gamma}_{\rm dip}$. Among others, it
depends on the total number of photons in and on the aperture of the
pulse \cite{palffy2020}, both experimental parameters which are not
exactly known at this time. The value of $\widetilde{\Gamma}_{\rm
  dip}$ serves as an input parameter for our calculation. We consider
values in the range $1$ -- $20$ MeV  and disregard any
  temperature dependence which could be the consequence of increased
  spreading widths for GDRs built up on highly excited
  states. According to the expression of $\widetilde{\Gamma}_{\rm
    dip}$ obtained in Ref.~\cite{palffy2020}, such a temperature
  dependence of $\Gamma^\downarrow$ would only slowly decrease the
  effective dipole width. Following Ref.~\cite{Pal15} we set
$\widetilde{\Gamma}_{\mathrm{dip}} = W^2 _{01;01} (i,1) \rho_1(0,1)$.
Photon absorption at excitation energy $k \hbar \omega_0$ by an
$m$p-$m$h state leading to an $m'$p-$m'$h state is then governed by
the effective absorption rate,
\begin{eqnarray}
W_{k k+1 ; m m'}^{2}(i,k+1) \rho_{m'}(i, k+1)\nonumber  \\
=W_{01 ; 01}^{2}(i,1) \rho^{\mathrm{acc}}_{mm'}(i, k+1) \ .
\end{eqnarray}
Here \(\rho^{\mathrm{acc}}_{m m'}(i,k+1)\) with $m' = m, m + 1$ is the
density of states in class $(i, k + 1, m')$ that are accessible from
class $(i, k, m)$. Using symmetry of the matrix elements we find for
stimulated dipole emission
\begin{eqnarray}
    &&W^2_{kk-1;mm'}(i,k-1) \rho_{m'}(i,k-1) \nonumber \\
  &&=W_{01 ; 01}^{2}(i,1)\rho^{\mathrm{acc}}_{m m'}(i, k)
  \frac{\rho_{m'}(i,k-1)}{\rho_m(i,k)} \ \, ,
\end{eqnarray}
with $m'=m, \, m - 1$. The densities are worked out in
Sec.~\ref{dens}.

\subsubsection{Neutron decay}

Neutron decay is described as an evaporation process for which we use
the Weisskopf estimate~\cite{Pal15}. Neutron decay of states in class
$(i, k, m)$ populates states in the daughter nucleus $(i + 1)$. The
latter cover a continuum of energies which extends from zero to \((k +
1 / 2) E_{L}-(V-F)\). Here $V - F$ is the neutron binding energy in
the shell model. As described in Sec.~\ref{basi}, the states are
grouped into classes $((i + 1), k', m')$, where $k' \hbar \omega_0$
ranges from zero to an upper bound given by $k \hbar \omega_0 - V +
F$.   Neutron evaporation from a target nucleus in class
  $m$p-$m$h leads to a state in class $(m-1)$p-$m$h in the daughter
  nucleus. As stated in Sec.~\ref{basi}
 we do not use such states in our calculation. Instead,
    we approximate neutron decay by considering only transitions with
    $m' = m$ and $m' = m - 1$. The rate for either
    transition is given by
\begin{eqnarray}
  &&\Gamma_{N}\left(i, k \to k^{\prime}, m\to m'\right) =\frac{1}{2 \pi
    \rho_{m}(i, k)} \nonumber \\
  &&   \times \int \limits_{\left(k^{\prime}-1 / 2\right) \hbar
    \omega_0}^{(k'+1 / 2) \hbar \omega_0} d E \, \rho_{m'}\left(i+1, E\right)
  \ .
 \label{gamma_n}
\end{eqnarray} 
The total rate for depletion of class $(i, k, m)$ is written as
\begin{eqnarray}
   && \Gamma_{N}(i, k, m) =\frac{1}{2 \pi  \rho_{m}(i, k)} \\
&&   \times \int \limits_{0}^{(k+1 / 2) \hbar \omega_0 -(V-F)} d E \
\sum_{m' = m, m - 1} A_{m'} \rho_{m'} \left(i+1, E\right) \ . \nonumber
\label{gamma_n1}
\end{eqnarray}
To avoid double counting we must have $A_m + A_{m - 1} = 1$. As shown below, the two terms in the summation in Eq.~(\ref{gamma_n}) are practically equal, and we choose $A_m = 1$, $A_{m - 1} = 0$ in what follows. The last term in
Eq.~(\ref{eq:me3}) must be modified accordingly. For the short chains
  $i = 0, 1, 2, \ldots, n$ that we actually consider, we simplify the
  calculations by keeping $V - F= 8$ MeV fixed. We thereby neglect the
  odd-even staggering of binding energies and level densities. These
  run in parallel and, therefore, largely compensate each other in the
  neutron decay widths.

\subsubsection{Level densities \label{dens}}

The level densities $\rho_m(i, k)$ are calculated using the method
developed in Ref.~\cite{Pal13b} for the total level density of
spin-zero states in nucleus $A$ as a function of excitation
energy. The calculation uses as input the single-particle level
density $\rho_1(\ve)$, a continuous function of energy $\ve$. In this
work we consider both an energy-independent function $\rho_1(\ve)$
which yields a constant spacing of single-particle levels of 0.88 MeV
(used for $A = 42$) and a linear energy dependence,
\be
\rho^{(1)}_1(\ve) = \frac{2 A}{F^2} \ve \ ,
\label{4}
\ee
that is approximately valid for $A = 100$. The single-particle
energies $\ve_j$ with $j = 1, 2, \ldots$ are obtained from
Eq.~(\ref{4}) via the condition $j = \int_0^{\ve_j} {\rm d} \ve'
\ \rho^{(1)}_1(\ve')$. We use $V = 45$ MeV and $F = 37$ MeV for all
nuclei in the neutron decay chain. These values determine the total
number of bound single-particle states \cite{Pal13b}. 
For $A = 42$, that number is $51$. For $A = 100$ and the linear
dependence of $\rho^{(1)}_1$ in Eq.~(\ref{4})), that number is $148$.

It was shown in Ref.~\cite{Pal13b} that when the number of nucleons is
large the method of calculation fails to properly describe the tails
of the level densities $\rho_m(i, k)$ at small excitation energies. In
that region we extrapolate the level densities. That is done for a
small fraction (typically approximately 10$\%$, for certain
particle-hole classes however up to 35$\%$) of the total relevant part
of the spectrum.

As in Ref.~\cite{Pal13b} the density $\rho^{\mathrm{acc}}_{mm'}(E)$ of
accessible states is calculated using the Fermi-gas model and
$\rho^{(1)}_1$ as given in Eq.~(\ref{4}). Here we sketch the
modifications that arise from the existence of particle-hole
classes. The Fermi distributions for holes and particles with
single-particle energy $\varepsilon$ are given, respectively, by
\begin{align}
  n_{A - m, E} (\varepsilon) & =\frac{\Theta(F-\varepsilon)}{1+\exp \{
    \beta \varepsilon + \alpha _{A-m}\}} \ , \nonumber \\
  \quad n_{m, E}(\varepsilon) & =\frac{\Theta(\varepsilon-F)}{1 +
    \exp \{ \beta \varepsilon + \alpha_{m}\}} \ .
\end{align} 
The first expression describes \(A-m\) particles below the Fermi level
$F$ (corresponding to the $m$ holes). The second expression describes
\(m\) particles above \(F\). Both expressions carry the same parameter
$\beta$ because particles and holes have the same temperature. The
parameters $\beta, \alpha_{A - m}, \alpha_m$ are determined by the
constraints
\begin{equation}
  \label{cons}
    \begin{aligned}
      A-m=&\int\limits_{0}^{F}  d {\varepsilon}\,  n_{A-m}(\varepsilon)
      \rho_{1}(\varepsilon) \ ,\\
      m=&\int\limits_{F}^{V} d {\varepsilon}\, n_{m}(\varepsilon)
      \rho_{1}(\varepsilon) \ ,\\
      E=&\int\limits_{0}^{V} d {\varepsilon}\, \varepsilon\left[n_{m}
      (\varepsilon)+n_{A-m}(\varepsilon)\right] \rho_{1}(\varepsilon) \ .
    \end{aligned}
\end{equation}
These impose fixed hole number, fixed particle number, and fixed total
energy $E$, respectively. 

\begin{figure}[ht]
\includegraphics[width=\linewidth]{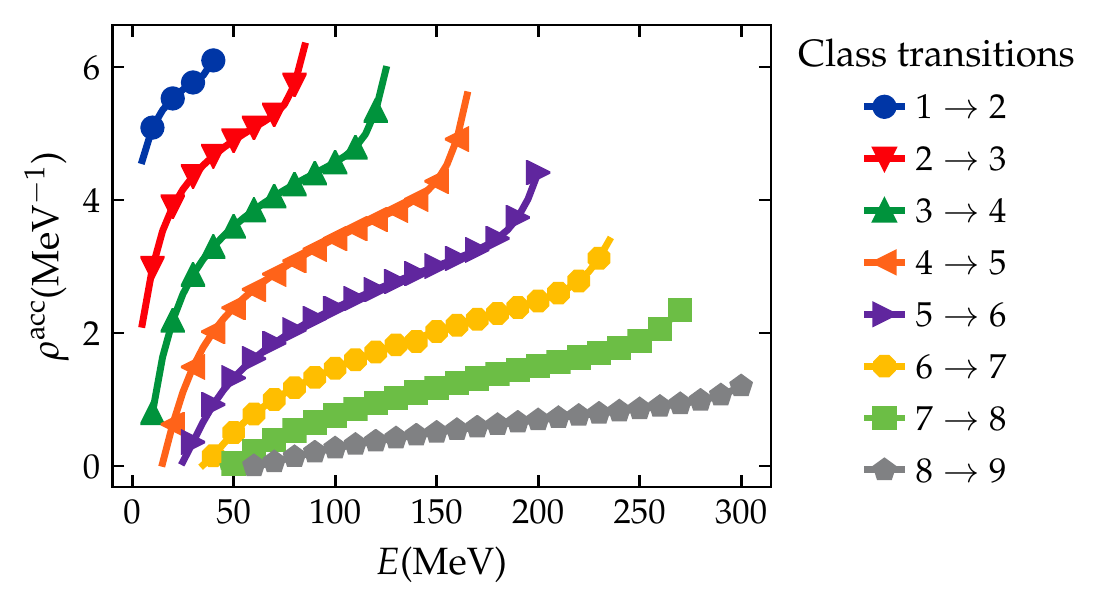}
    \caption{Density of accessible states $\rho^{\text{acc}}_{m m+1}
      (E)$ for case (i) (see text) in the constant-spacing model as a
      function of excitation energy \(E\) for $A=42$ particles, $51$
      single-particle states, $V =45$ MeV, $F = 37$ MeV, and $\hbar
      \omega_0 = 5$ MeV. Colors and symbols correspond to different transitions between particle-hole
      classes $m\rightarrow m+1$.}
    \label{fig:acc-di-densities-1}
\end{figure}

The absorption of a photon of energy $\hbar \omega_0$ involves either
one of two processes: (i) A nucleon absorbs the energy $\hbar
\omega_0$ and is thereby promoted from a single-particle state below
$F$ to a single-particle state above $F$ (without being promoted to
the continuum). That causes a transition from class $(i, k, m)$ to
class $(i, k + 1, m + 1)$. (ii) A particle absorbs the energy $\hbar
\omega_0$ (without being promoted to the continuum), or a hole absorbs
$\hbar \omega_0$ without exceeding the Fermi energy. That causes a
transition from class $(i, k, m)$ to class $(i, k + 1, m)$.

For case (i), the energy $\ve$ of the nucleon prior to photon
absorption must obey \(F- \hbar \omega_0 < \varepsilon < F\). The
probability of finding an occupied single-particle state at energy
$\varepsilon$ below $F$ is $n_{A-m, E}(\varepsilon)$, and the
probability of finding an empty single-particle state with energy
$\varepsilon + \hbar \omega_0 > F$ is $(1-n_{m, E}(\varepsilon + \hbar
\omega_0))$. The density of accessible states is, thus, given by
\begin{eqnarray}
\label{eq:rhoaccph-1}
\rho^{\text{acc}}_{m m+1} (E)& =& \int \limits ^{F} _{F -
  \hbar \omega_0} d {\varepsilon}\, n_{A - m, E}(\varepsilon) [1-n_{m,
    E}(\varepsilon + \hbar \omega_0)] \nonumber \\ &\times&
\rho_1(\varepsilon) \rho_1(\varepsilon + \hbar \omega_0) \ .
\end{eqnarray}
Figure~\ref{fig:acc-di-densities-1} shows $\rho^{\text{acc}}_{m m+1}(E)$
versus energy $E$ for several values of $m$ and for parameters given
in the figure caption. The density of accessible states is a
monotonically increasing function of energy for all particle-hole
classes. It decreases with increasing $m$ of the particle-hole class.

\begin{figure}[ht]
\includegraphics[width=\linewidth]{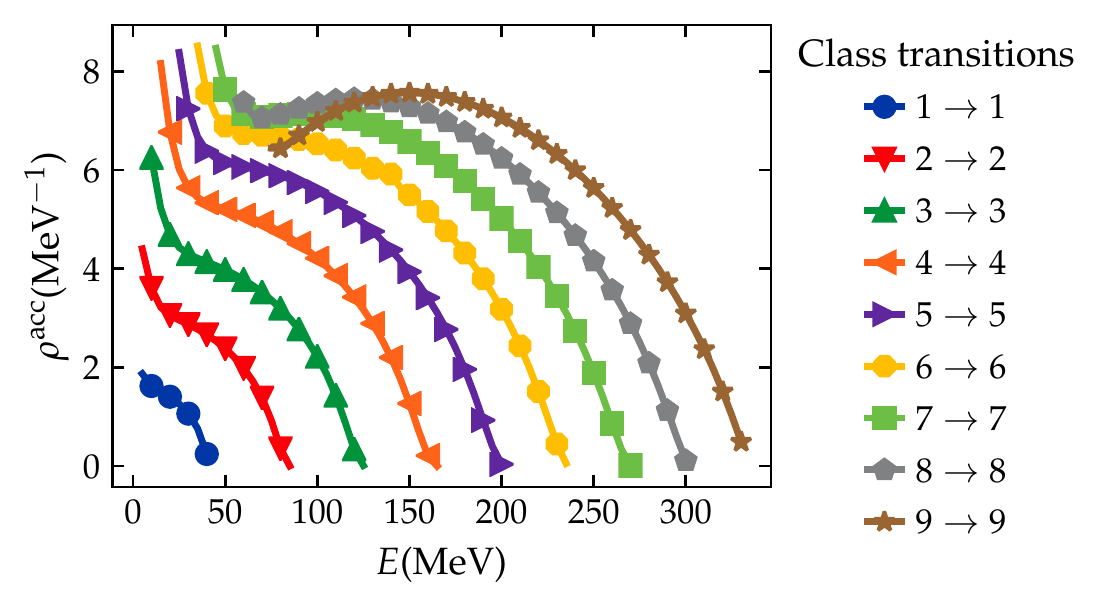}
    \caption{Density of accessible states $\rho^{\text{acc}}_{m m}
      (E)$ for case (ii) (see text) in the constant spacing model as a
      function of excitation energy \(E\). We use the same parameters
      as for Fig.~\ref{fig:acc-di-densities-1}. Colors and symbols correspond to
      transitions within different particle-hole classes $m$.}
    \label{fig:acc-di-densities-2}
\end{figure}

For case (ii) we obtain analogously
\begin{eqnarray}
\label{eq:rhoaccph-2}
  \rho^{\text{acc}}_{mm} (E)&=& 
  \int \limits _{0} ^{F - \hbar \omega_0} d {\varepsilon} \,
  n_{A - m, E}(\varepsilon) [1-n_{A - m,E}(\varepsilon + \hbar \omega_0)]
  \nonumber \\
  &\times& \rho_1(\varepsilon) \rho_1(\varepsilon + \hbar \omega_0)
    \nonumber \\ 
    &+& \int \limits _{F} ^{V - \hbar \omega_0} d {\varepsilon} \,
    n_{m, E}(\varepsilon) [1 - n_{m, E}(\varepsilon + \hbar \omega_0)]
    \nonumber \\
  & \times& \rho_1(\varepsilon) \rho_1(\varepsilon + \hbar \omega_0)\ .
\end{eqnarray}
For the same set of parameters as used in
Fig.~\ref{fig:acc-di-densities-1}, Fig.~\ref{fig:acc-di-densities-2}
shows results for $\rho^{\text{acc}}_{mm}(E)$ versus $E$ for several
values of $m$. For large $E$ that function decreases with increasing
$E$. For such energies, the sparseness of empty levels makes it
increasingly difficult to add the energy $\hbar \omega_0$ to a
particle or a hole. That effect is absent in $\rho^{\text{acc}}_{m m +
  1}(E)$.

\section{Numerical Results \label{numres}}

We calculate the time-dependent occupation probabilities $P_m(i, k,
t)$ for two medium-weight target nuclei, $A = 42$ and $A = 100$, that
interact with a short pulse of MeV photons. These nuclei are taken to
be generic for  their range of mass values. We solve
Eq.~(\ref{eq:me3}) numerically for several choices of the effective
dipole width $\widetilde{ \Gamma}_{\rm dip}$ and of the length $(n +
1)$ of the decay chain. Equation~(\ref{eq:me3}) is written in matrix form
as $\dot{P} = \mathcal{M} P$. The elements of the column vector \(P\)
are the occupation probabilities $P_m(i, k, t)$ labeled by an overall
index $j$ covering the set $(i, k, m)$. The initial condition $P_m(i,
k, 0) = \delta_{i, 0} \delta_{k, 0} \delta_{m, 0}$ mimics the ground
state of the target nucleus.   The matrix $\mathcal{M}$ is independent
of time. The number of elements in $\mathcal{M}$ ranges from \(670\)
for \(A=42\) to \(\sim 10000\) for \(A=100\) per neutron decay
generation. These elements vary over 8 orders of magnitude for nuclei
with mass number \(A=42\), and over 70 orders of magnitude for nuclei
with mass number \(A=100\). Diagonalization of this matrix, therefore,
poses a stiff problem. As in Ref.~\cite{Pal15} we treat the extremely
stiff differential equations~(\ref{eq:me3}) via a matrix exponential
method. We use the Chebyshev rational approximation method (CRAM)
which is known for its success in solving burnup
equations~\cite{Pusa2011,Pusa2016}. For all the calculations we use
CRAM with partial fraction decomposition and an approximation of order
20~\cite{Pusa2011}.  Despite the efficiency of CRAM, the size and stiffness of the 
matrix $\mathcal{M}$ restricts our present calculations to nuclear mass numbers $A\le 100$.

Our numerical calculations yield values for the occupation probability
$P_m(i, k, m)$. In the contour plots for $P_m(i, k, m)$ we convert $k$
to energy $E$ via $E = k \hbar \omega_0$.  We use $\hbar \omega_0 = 5$
MeV throughout. This value lies well within the planned range
of the Extreme Light Infrastructure \cite{ELI-web} and the Gamma Factory \cite{gamma-fact} facilities mentioned in the Introduction.  

\subsection{Light medium-weight nuclei ($A=42$)} 
\label{sec:constant_spacing}

\begin{figure}[htbp]
    \centering
    \includegraphics[width=0.95\linewidth]{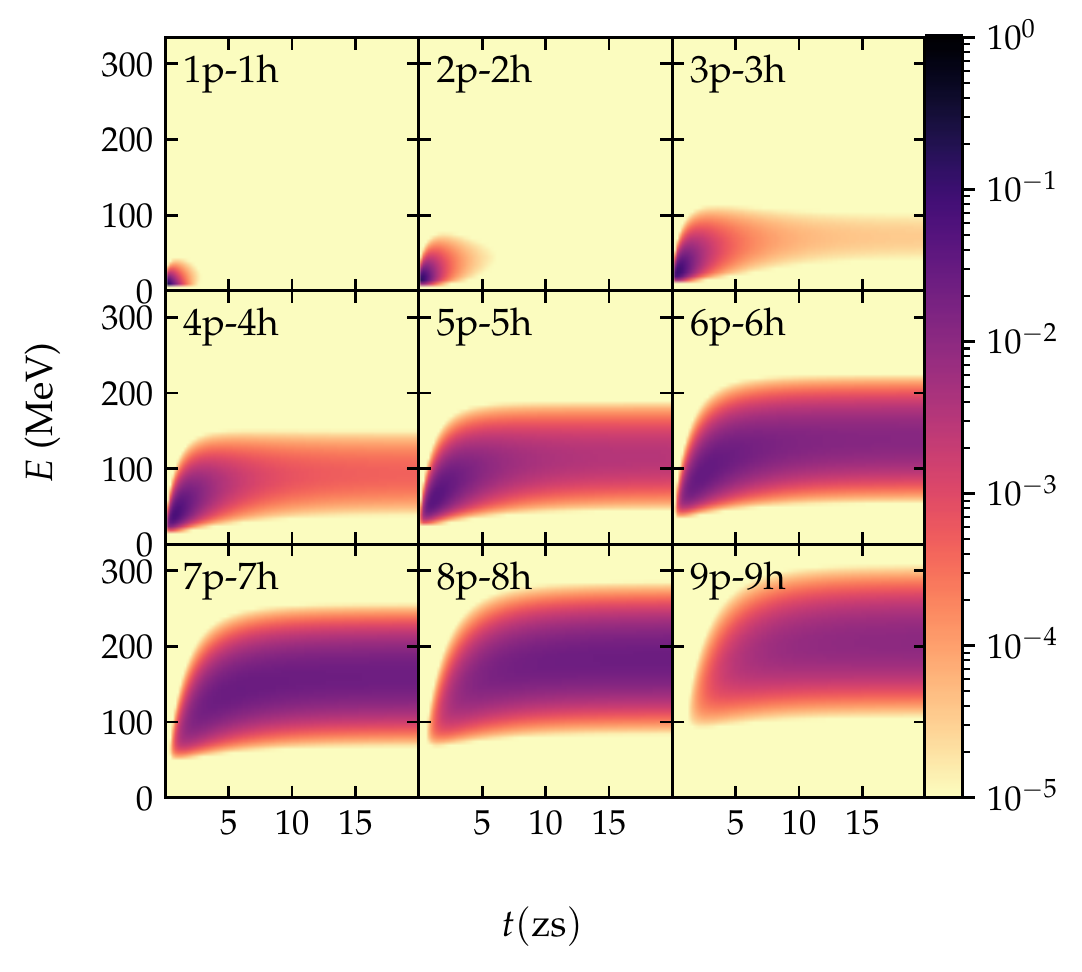}
    \caption[OP, p-h classes, constant spacing, \(A=42\)]{Contour
      plots of the time-dependent occupation probabilities $P_m(0, k,
      t)$ of a light medium-weight nucleus with $A=42$ as a function of
      excitation energy $E$ for the accessible particle-hole classes. Neutron evaporation is not included. The
      parameters used are given in the text.}
    \label{fig:classesA42}
\end{figure}

We first consider the comparatively simple case of small nucleon
number $A = 42$ and constant level density \( \rho_1(\varepsilon)
\equiv\mathrm{const} = A/F \) with $51$ single-particle states.
Figure~\ref{fig:classesA42} shows the occupation probabilities versus
time and excitation energy for the target nucleus in the absence of
neutron decay ($\Gamma_N = 0$) for \(\widetilde{
  \Gamma}_{\mathrm{dip}} = 5 \)~MeV and a duration time of the laser
pulse \( 1 / \sigma = 20\)~zs. During the process, particle-hole
classes up to $m = 9$ are populated, with the higher $m$-values
requiring larger excitation energy.

\begin{figure}[htbp]
    \centering
    \includegraphics[width=\linewidth]{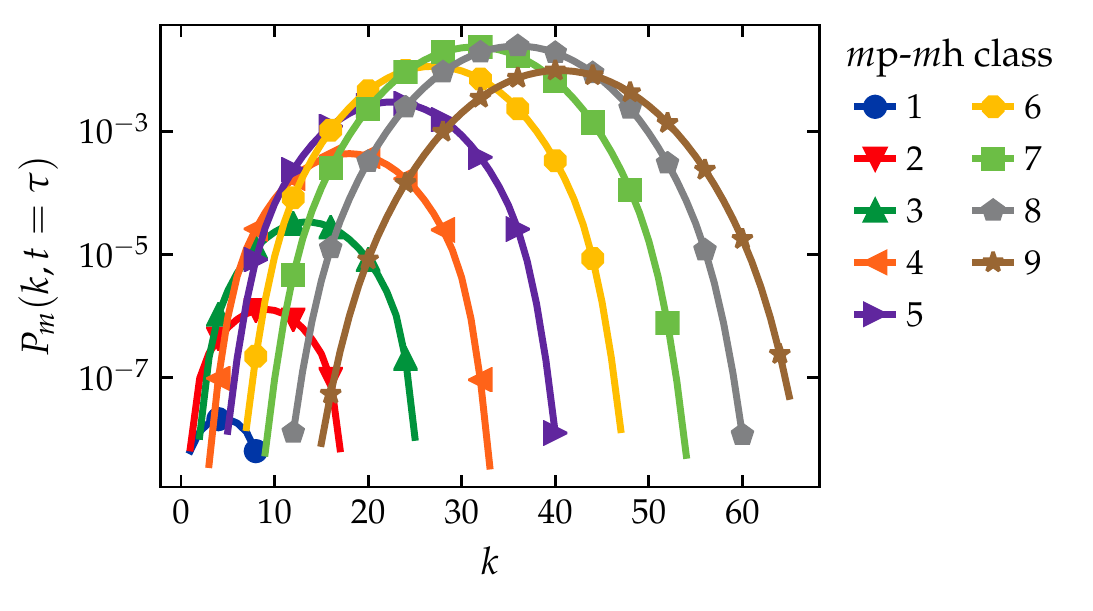}
    \caption[OP at laser pulse termination, constant spacing,
      \(A=42\)]{A cut through the contour plots in
      Fig.~\ref{fig:classesA42} at the termination \(t = 20\)~zs
      of the laser pulse. The occupation probabilities are shown as
      functions of the number of absorbed photons \(k\) for different
      particle-hole classes.  }
    \label{fig:saturprob.a42ph}
\end{figure}

A cut in the contour spectra of Fig.~\ref{fig:classesA42} at $t =
20$~zs (i.e., at the end of the laser-nucleus interaction) is shown in
Fig.~\ref{fig:saturprob.a42ph}. For each class $m$ of particle-hole
numbers, the occupation probabilities display a maximum. It occurs at
the energy for which the rates of dipole absorption and stimulated
dipole emission are equal. There the particle-hole level density
$\rho_m(0, k)$ has its maximum. Beyond that peak, stimulated dipole
emission outweighs dipole absorption, the excitation process saturates,
and further excitation becomes increasingly unlikely. Inspection of
Fig.~\ref{fig:classesA42} shows that prior to termination of the pulse
at $t = 20$~zs, classes $m = 5$ and $m = 6$ are closer to saturation
(the occupation probabilities run parallel to the abscissa) than
classes with higher values of $m$. We conclude that for different
classes saturation is achieved at different times, and the system as a
whole is saturated when the ``slowest'' class is saturated.
 
\begin{figure}[htbp]
    \centering
    \includegraphics[width=0.8\linewidth]{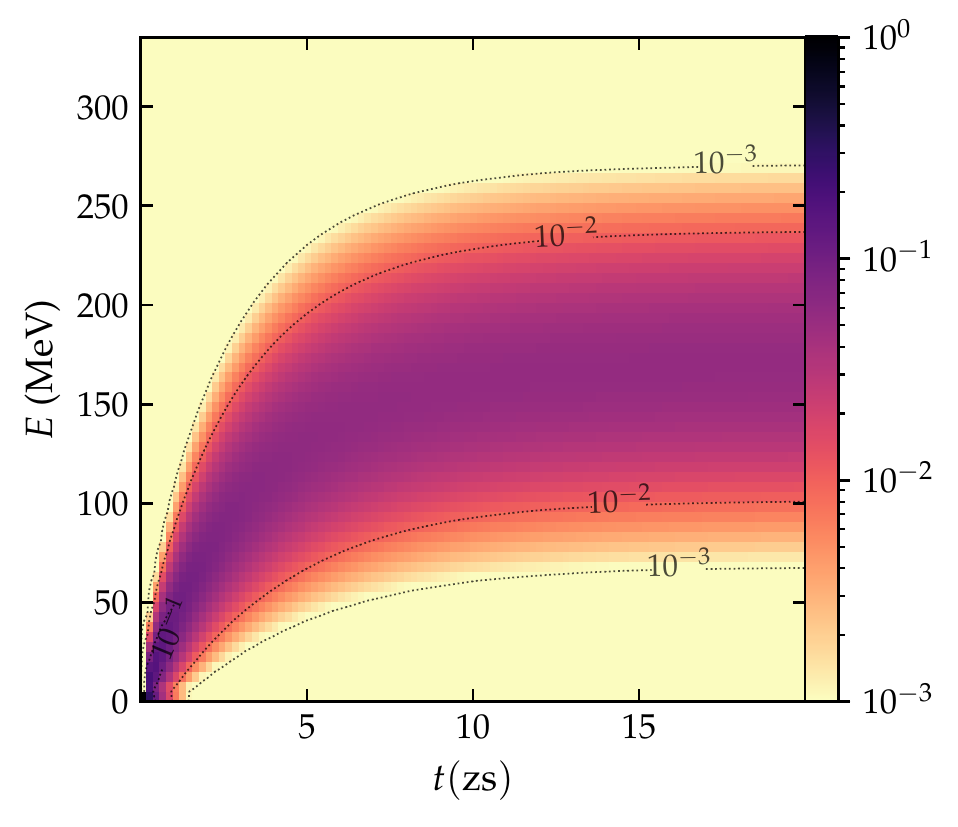}
    \caption[A=42, constant spacing, summed occupation
      probabilities]{Contour plot of the time-dependent total
      occupation probability as a function of excitation energy $E$
      and time $t$ for dipole absorption and stimulated emission
      only (no neutron decay). We use the same parameters as in
      Fig.~\ref{fig:classesA42} (see text).}
    \label{fig:summedprob.a42ph}
\end{figure}

Figure~\ref{fig:summedprob.a42ph} shows the total occupation probability
$P(0, k, t)=\sum_m P_m(0, k, t)$ of the target nucleus. Qualitatively,
that plot is similar to the quasiadiabatic results in
Ref.~\cite{Pal15}.

\subsection{Medium-weight nuclei $(A=100$)} 
\label{sec:medium_weight_nuclei}

For mass number $A = 100$ we use the single-particle level density in
Eq.~\eqref{4}. With a depth \(V= 45\)~MeV of the single-particle
potential, that gives a total of $148$ single-particle states.  The
increase in both particle number and number of states causes a
significant increase in the dimension of the matrix ${\cal M}$.
Fortunately, many matrix elements are zero and one can exploit that
sparseness to reduce the matrix dimension. The resulting matrix for
the target nucleus has dimension 5328. The resulting matrix for $n =
4$ generations of nuclei has dimension 21088.

Our calculations were performed for pulse durations of 40~zs and for
four choices of the effective dipole width, $\widetilde{\Gamma}_{\rm
  dip} = $1, 5, 10, and 20 MeV. We focus attention on
$\widetilde{\Gamma}_{\rm dip} = 20$~MeV (typical for the sudden
regime) and on $\widetilde{\Gamma}_{\rm dip} = 5$~MeV (relevant for a
comparison with results for the quasiadiabatic regime in
Ref.~\cite{Pal15}).

For $A = 100$, particle-hole classes up to $m = 48$ can be reached.
For a better understanding of our results we first show in
Fig.~\ref{rho_A100} the density of states versus $k$ for every fourth
particle-hole class. The excitation energy is given by the number $k$
of absorbed photons, each with energy $\hbar \omega_0 = 5$
MeV. Particle-hole classes with small (large) $m$ dominate at small
(large) energies, respectively. Densities with $m \approx 30$ have the
largest values. The total density (summed over all particle-hole
classes) is the envelope to these curves, with a maximum at $\approx
575$ MeV.

\subsubsection{No Neutron Evaporation}

\begin{figure}[htbp]
    \centering
    \includegraphics[width=\linewidth]{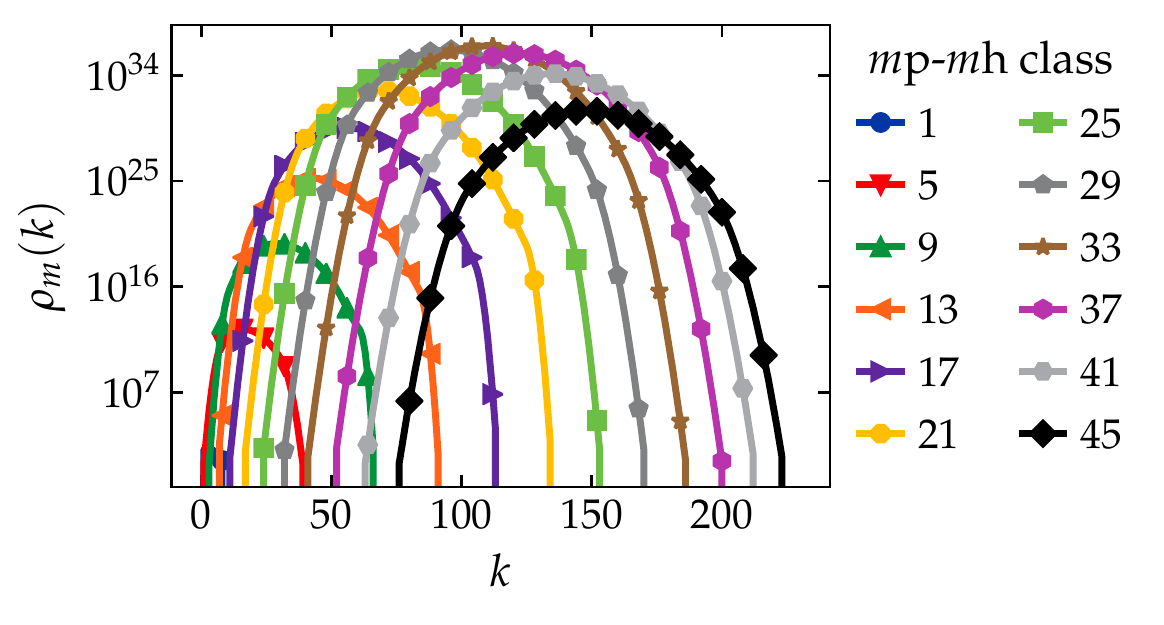}
    \caption[A=100, medium-weight, level densities]{Level densities
      $\rho_m(0, k)$ for a medium-weight nucleus $A=100$ as
      functions of excitation energy (measured in units of the
      number $k$ of absorbed photons). For clarity of illustration we present the
      densities only for every fourth particle-hole class. }
    \label{rho_A100}
\end{figure}

\begin{figure}[htbp]
    \centering
    \includegraphics[width=\linewidth]{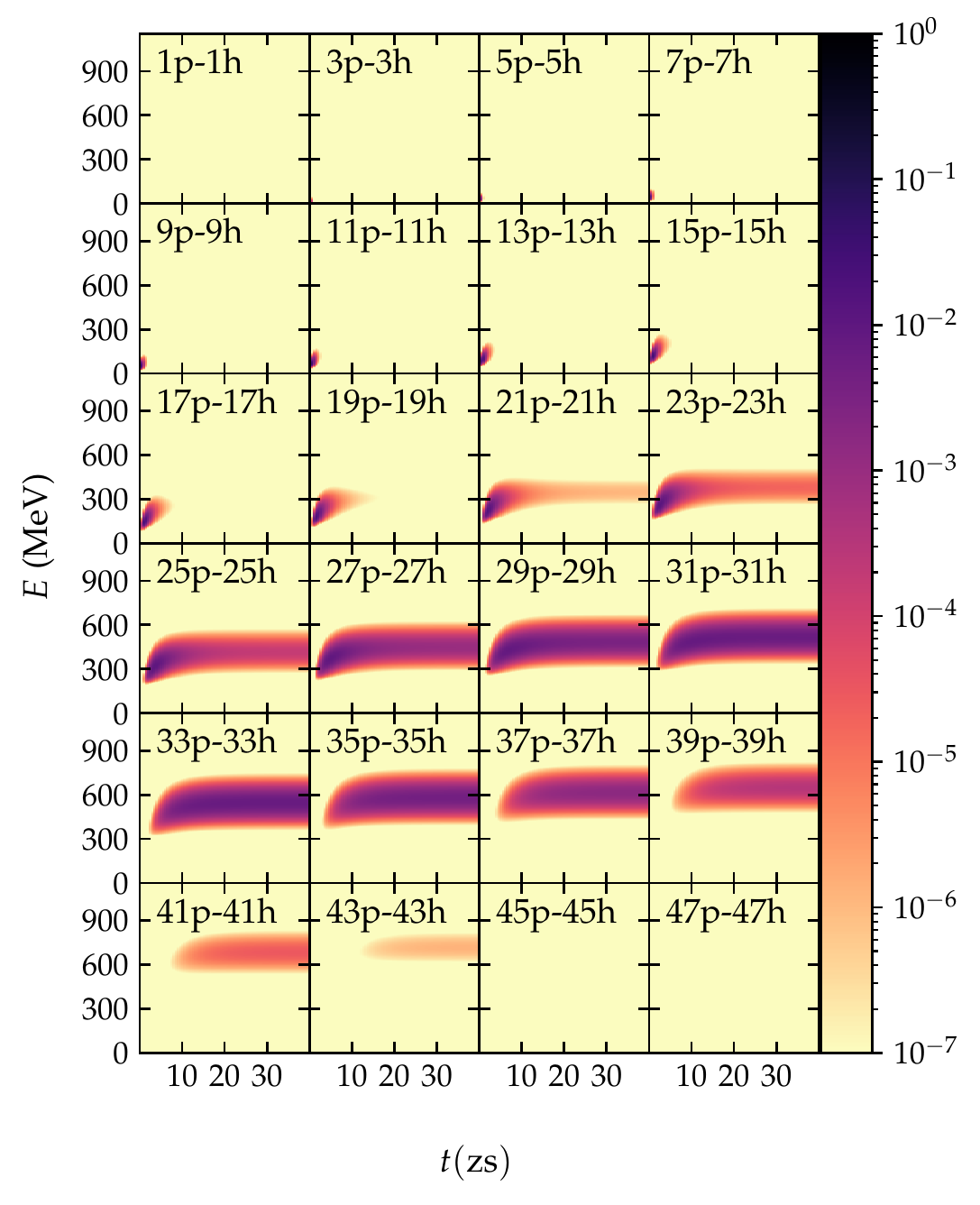}
    \caption[A=100, medium-weight, occupation probabilities for p-h
      classes]{Contour plots of the time-dependent occupation
      probabilities $P_m(0, k, t)$ for every second accessible
      particle-hole class from \(1\)p-\(1\)h to \(47\)p-\(47\)h as
      functions of excitation energy $E$ for dipole absorption and
      stimulated emission only (no neutron emission). The parameters
      are mass number \(A=100\), the single-particle level density as
      given in Eq.~\eqref{4}, $\widetilde{\Gamma}_{\mathrm{dip}} = 20
      \)~MeV, and \(\tau = 40\)~zs. }
    \label{fig:classes.a100}
\end{figure}

We first focus attention on the time evolution of the target nucleus,
disregarding neutron evaporation. Fig.~\ref{fig:classes.a100}
presents the occupation probabilities as functions of time and
excitation energy for $\widetilde{\Gamma}_{\rm dip} = 20$~MeV. To be able to display the populations of low $m$ and high $m$
particle-hole classes in the same plot, the scale ranges from $10^{-
  7}$ to unity and comprises two orders of magnitude more than the
plot of Fig.~\ref{fig:classesA42}. Classes with small numbers of
particle-hole pairs are populated in the first stages of
photoexcitation. Classes with $m$ between \(23\) and \(37\) are then
occupied rapidly and stay populated until the end of the laser
pulse. The classes with the highest particle-hole numbers are
populated poorly and only late when sufficient energy was
transferred to reach the domain of excitation energy where their
densities are large. Figure~\ref{rho_A100} shows that these densities
reach their maxima at an energy higher than the saturation energy,
that these maxima are lower than those of the middle classes, and that
these maxima strongly decrease with increasing $m$ beyond the 36p-36h
class. That is in contrast to the case of constant spacing;
cf.~Fig.~\ref{fig:classesA42} where the maximum of the level density
for the highest 9p-9h class is not significantly smaller than that of
the neighboring 8p-8h class which dominates all the other classes.

\begin{figure}[htbp]
    \centering
   \includegraphics[width=\linewidth]{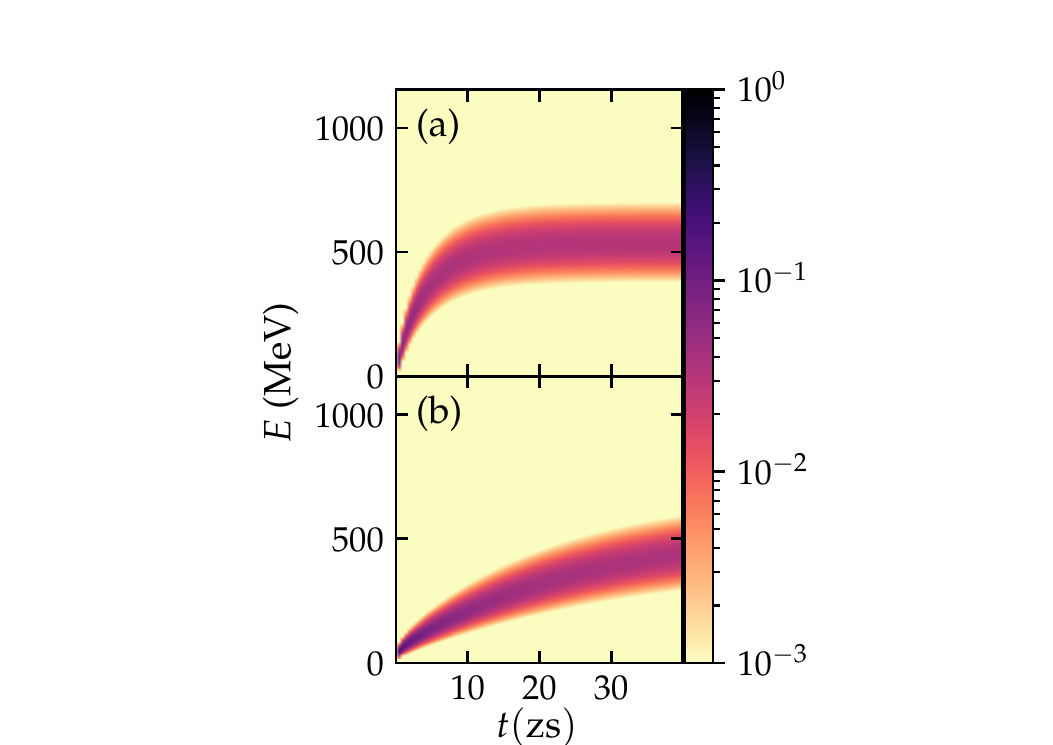}
    \caption[A=100, medium-weight, occupation probabilities for p-h
      classes]{Contour plots of the time-dependent total occupation
      probabilities $P(0, k, t)$. (a) Photon absorption or emission
      allows for the two possible processes presented in Figs.~\ref{w} and
      \ref{ww}. (b) Photon absorption or emission allows only for the
      process in Fig.~\ref{w}. The parameters are the same as in 
      Fig.~\ref{fig:classes.a100}. }
    \label{fig:comp2chan}
\end{figure}

Figure \ref{fig:classes.a100} shows that all classes reach
saturation. The same is true for the total occupation probability
$P(0, k, t)=\sum_m P_m(0, k, t)$ shown in Fig.~\ref{fig:comp2chan}(a).
Saturation is reached at $t\approx20$~zs, and $P(0, k, t)$ remains
constant thereafter. We note the qualitative similarity with results
obtained in Ref.~\cite{Pal15} for the quasiadiabatic regime. It is
interesting to note that the total occupation probability is sensitive
to the mechanism of photoabsorption. Indeed, allowing only for the
processes described in Fig.~\ref{w} (change of particle-hole class
for each photon absorption or emission process), the time scale for
excitation increases dramatically. Figure~\ref{fig:comp2chan}(b) shows
the total occupation probability calculated using $m'=m\pm 1$ only in
the dipole excitation and emission part of Eq.~(\ref{eq:me3}) for
\(\widetilde{ \Gamma}_{\mathrm{dip}}=20\)~MeV. Comparison with
Fig.~\ref{fig:comp2chan}(a) shows the increase of the time scale for
photoexcitation. Saturation is not yet reached even at $t=40$~zs.

\begin{figure}[htbp]
    \centering
    \includegraphics[width=\linewidth]{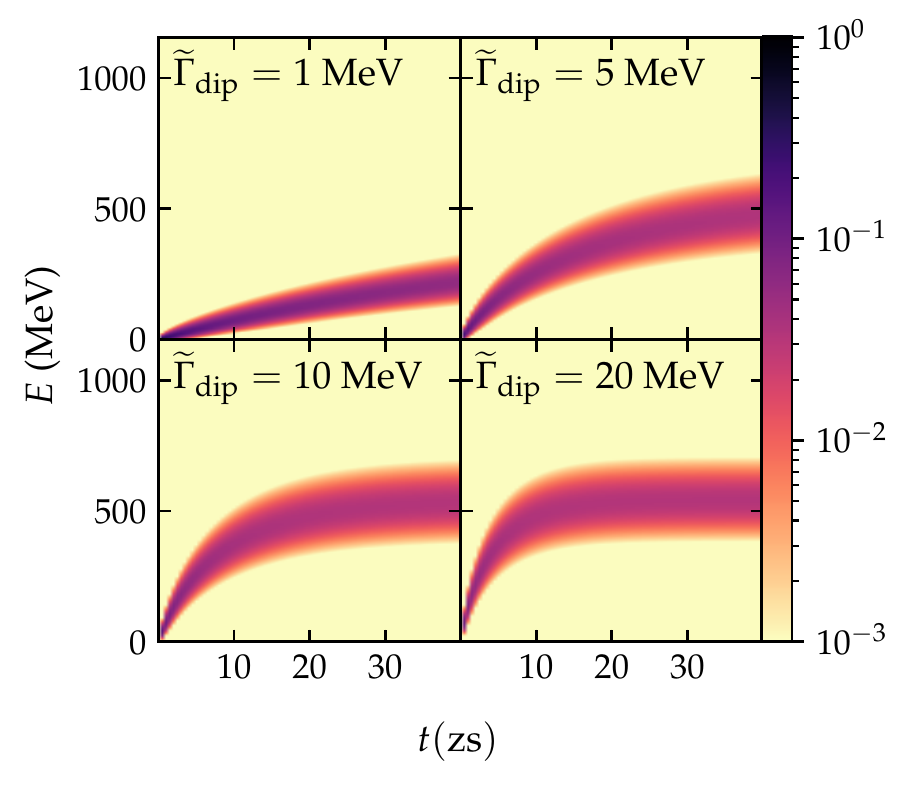}
    \caption[A=100, OP different \(N \Gamma_{\mathrm{dip}}\)]{Contour
      plots of the total time-dependent occupation probabilities $P(0,
      k, t)$ versus time $t$ and excitation energy $E$ for four values
      of \(\widetilde{ \Gamma}_{\mathrm{dip}}\) as indicated. The
      other parameters are the same as in
      Fig.~\ref{fig:classes.a100}. }
    \label{fig:dips.a100}
\end{figure}

Figure~\ref{fig:dips.a100} presents contour plots of the total
occupation probabilities for four choices of the effective dipole
width, \(\widetilde{\Gamma}_{\mathrm{dip}}=\) 1, 5, 10, and 20
MeV. The comparison shows whether and how quickly saturation is
reached. For \(\widetilde{ \Gamma}_{\mathrm{dip}}=1\)~MeV and \(
\widetilde{\Gamma}_{\mathrm{dip}}=5\)~MeV saturation requires duration
times longer than the pulse duration 40~zs used. For \(
\widetilde{\Gamma}_{\mathrm{dip}}=10\)~MeV saturation is reached at 40
zs, and for \(\widetilde{\Gamma}_{\mathrm{dip}}=20\)~MeV, it is not
possible to further transfer energy into the nucleus for times $t \geq
20$~zs.  Comparing with the case of $A=42$,
\(\widetilde{\Gamma}_{\mathrm{dip}} = 5\) MeV, and constant
single-particle level spacing shown in
Fig.~\ref{fig:summedprob.a42ph}, we notice that for $A=100$ the same
effective dipole rate $\widetilde{\Gamma}_{\mathrm{dip}} = 5\)~MeV
does not bring the nucleus to the same degree of saturation at
$t=20$~zs. Saturation requires either a longer duration time of the
laser pulse or a greater effective dipole rate. That is because
saturation in the nucleus \(A=100\) occurs at a substantially higher
energy. For $A = 42$ (for $A = 100$), the maximum of the total level
density is at \(165\) MeV (at \(533\) MeV, respectively). [For $A =
100$ and for the single-particle level density as given in
Eq.~\eqref{4}, the total level density is a slightly asymmetric
function of energy].

\begin{figure}[htbp]
    \centering
    \includegraphics[width=\linewidth]{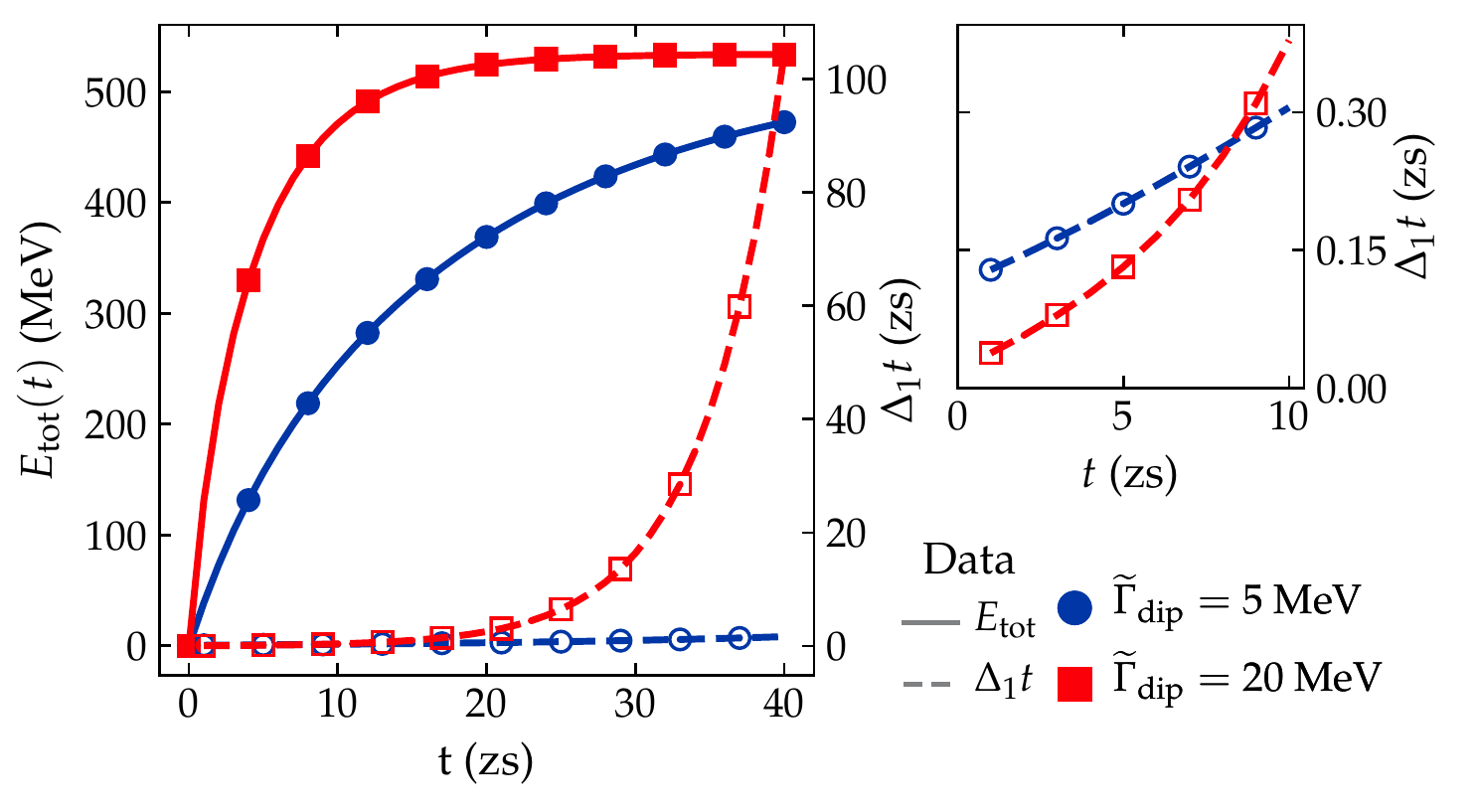}
    \caption[A=100, Etot]{Total excitation energy $E_{\rm tot}$ (solid
      lines, left vertical axis) and approximate time interval between
      two photon absorptions $\Delta_1t$ (dashed lines, right vertical
      axis) as a function of time for \(\widetilde{
        \Gamma}_{\mathrm{dip}}=5\)~MeV (blue circles) and \(\widetilde{
        \Gamma}_{\mathrm{dip}}=20\)~MeV (red squares). The inset zooms in the
      interval of the first 10 zs for $\Delta_1t$.}
    \label{Etot}
\end{figure}

To compare results for different effective dipole widths we use the
total excitation energy $E_{\rm tot}(t) = \hbar \omega_0 \sum_{k, m} k
P_m(0, k, t)$ of the nucleus at time $t$, and the time interval
$\Delta_1 t(E) = t(E+\hbar\omega_0)-t(E)$, where $t(E)$ is the
earliest time at which the total energy $E$ is reached. At very short
times, $E_{\rm tot}(t)$ grows linearly with time $t$, and $\Delta_1t$
is a measure of the time interval between the successive absorption of
two photons. With increasing excitation energy stimulated emission
becomes important, $E_{\rm tot}(t)$ grows less strongly than linearly,
and $\Delta_1t$ increases correspondingly. Figure~\ref{Etot} shows
$E_{\rm tot}$ and $\Delta_1t$ as functions of time for \(\widetilde{
  \Gamma}_{\mathrm{dip}}=5\)~MeV and \(\widetilde{
  \Gamma}_{\mathrm{dip}}=20\)~MeV. As expected, at short times $E_{\rm
  tot}(t)$ increases significantly faster for the bigger of the two
effective dipole rates. The time between two successive photon
absorption processes is correspondingly shorter and, thus, more
competitive with the nuclear relaxation time. We have to keep in mind,
however, that the relaxation time itself becomes shorter, too, with
increasing excitation energy. Moreover, Fig.~\ref{Etot} shows that
induced photon emission and, eventually, saturation become important
very early for \(\widetilde{ \Gamma}_{\mathrm{dip}}=20\)~MeV, so that
$\Delta_1 t$ ceases to be a measure of the time interval between
successive photon absorption processes.

\subsubsection{Comparison with the quasiadiabatic case}

\begin{figure}[htbp]
    \centering
    \includegraphics[width=\linewidth]{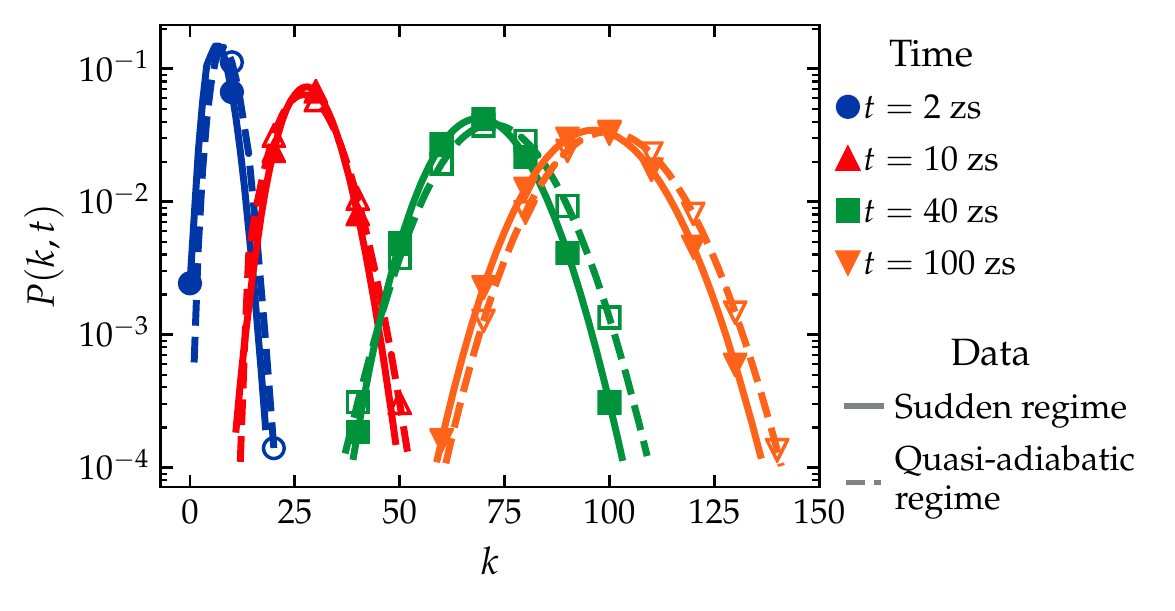}
    \caption[A=100, comparison of occupation probabilities, 5
      MeV]{Total occupation probabilities (solid lines) compared with
      results from Ref.~\cite{Pal15} for the quasiadiabatic regime
      (dashed lines) as a function of the number $k$ of absorbed
      photons at times $t = $ 2, 10, 40, and 100 zs and for
      \(\widetilde{\Gamma}_{\mathrm{dip}} = 5\)~MeV.}
    \label{fig:comp.5G.a100}
\end{figure}

In the quasiadiabatic regime, equilibration is instantaneous. Only
total level densities and total occupation probabilities enter the
calculation. If in the sudden regime the dipole width \(\widetilde{
  \Gamma}_{\mathrm{dip}} \) is so small that equilibration happens
between each pair of subsequent photon processes, our values for the
total occupation probabilities must agree with the ones calculated for
the quasiadiabatic regime. In Fig.~\ref{fig:comp.5G.a100} we compare
our total occupation probabilities for \(\widetilde{
  \Gamma}_{\mathrm{dip}}=5\)~MeV and 100-zs pulse duration with
results from Ref.~\cite{Pal15}. The parameters are the same as in our
case with one exception. The dipole matrix element used in the
quasiadiabatic calculation differs from that of the present
calculation by a factor 2.3 because its definition involves a
different 1p-1h density $\rho_1(0,1)$; see
Ref.~\cite{Obloinsk1986}. In Fig.~\ref{fig:comp.5G.a100} we have
accounted for that difference. The figure shows snapshots of the
occupation probability at four time instants, $t=2,\, 10,\, 40$, and
$100$ zs. The results display good agreement. Slight differences in
the wings can be attributed to the different strategies to calculate
the tails of the level densities used here and in Ref.~\cite{Pal15}.
The two calculations agree only if in our calculation photon
absorption and emission both without and with change of particle-hole
class as shown in Figs.~\ref{w} and \ref{ww} are included.

\begin{figure}[htbp]
    \centering
    \includegraphics[width=\linewidth]{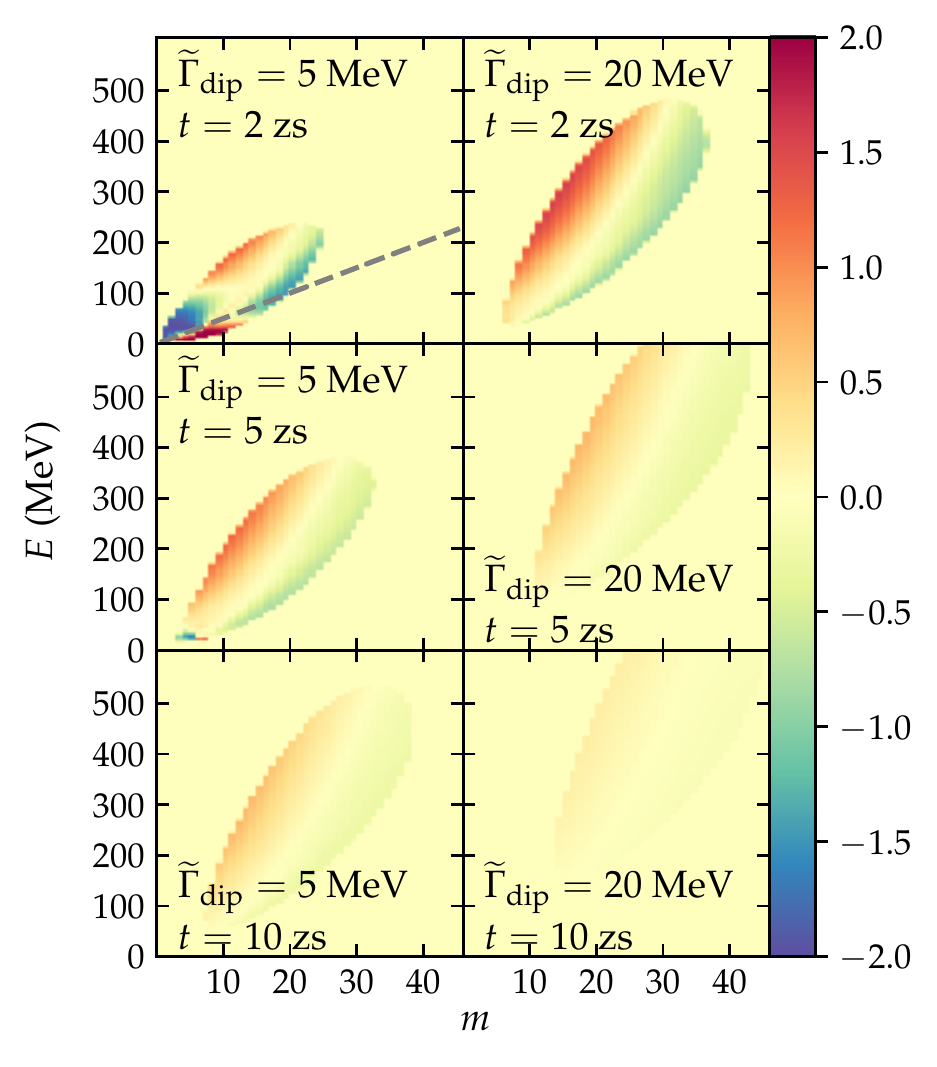}
    \caption[A=100, reldiff eq vs non-eq]{Relative difference $R_m(k)
      = 2[P^{\rm eq}_m(0, k, t) - P_m(0, k, t)]/[P^{\rm eq}_m(0, k, t)
        +P_m(0, k, t)]$ as a function of excitation energy $E$ and
      particle-hole class $m$. We consider snapshots at $t = 5, \, 10$
      and $20$ zs for effective dipole absorption rates
      $\widetilde{\Gamma}_{\mathrm{dip}} = 5 \)~MeV and $20$ MeV. The
      dashed line shows the function $E(m)=m\hbar\omega_0$ with
      $\hbar \omega_0 = 5$~MeV. See text for further
      explanations. }
    \label{fig:reldiff}
\end{figure}

For the generic value $\Gamma^\downarrow = 5$ MeV of the spreading
width for medium-weight nuclei, the nuclear relaxation time would be $
1 / \Gamma^{\downarrow}\simeq 0.13$ zs. However, our calculations
using the rates of Sec.~\ref{eq-sec} show that the relaxation time
actually depends on energy and particle-hole configuration and ranges
from less than 1 zs to a few zs. That indicates that for short times
(i.e., at the beginning of the laser pulse), photon absorption is
faster than equilibration. As a test, we compare in
Fig.~\ref{fig:reldiff} the occupation probabilities $P_m(0, k, t)$
obtained in our calculation with equilibrium values $P^{\rm eq}_m(0,
k, t) = \rho_m(0, k) P(0, k, t) / \rho(0, k)$. Here $\rho(0, k)$ is
the total level density and $P(0, k, t)$ is the total occupation
probability at energy $k$. We do so for
\(\widetilde{\Gamma}_{\mathrm{dip}} = 5\) and $20$~MeV, for three
instants of time ($t = 2, \, 5$, and $10$ zs), and for a range of
$k$ values (or excitation energies) and $m$ values. We display the
relative difference $R_m(k) = 2[P^{\rm eq}_m(0, k, t) - P_m(0, k,
  t)]/[P^{\rm eq}_m(0, k, t) + P_m(0, k, t)]$ in a contour plot. With
increasing excitation energy $E$ (or increasing $k$), the total number
of classes $(0, k, m)$ increases strongly whereas each absorbed photon
creates at most one additional particle-hole pair. Therefore we expect
that for fixed $k$ and prior to equilibrium, classes with small $m$
(large $m$) are overpopulated (underpopulated), corresponding to
$R_m(k) < 0$ ($R_m(k) > 0$, respectively). That expectation is
actually met for \(\widetilde{\Gamma}_{\mathrm{dip}} = 5\)~MeV, small
time $t = 2$ zs (upper left panel of Fig.~\ref{fig:reldiff}) and for
excitation energies up to $\approx 100$ MeV. The dashed line
corresponds to a process where each photon of energy $\hbar \omega_0 =
5$ MeV generates an additional particle-hole pair. A similar pattern
can be observed also for $t\simeq 0.5$~zs for
\(\widetilde{\Gamma}_{\mathrm{dip}} = 20\)~MeV, though not displayed
in Fig.~\ref{fig:reldiff}.

At first surprisingly, for all other data shown in the figure our
expectation fails, and the pattern is actually reversed. The
occupation probabilities at fixed excitation energy are largest for
classes with large particle-hole numbers. The behavior of the density
of accessible states $\rho^{\text{acc}}_{m m} (E)$ in
Fig.~\ref{fig:acc-di-densities-2} explains why this happens. The
densities of accessible states $\rho^{\text{acc}}_{m m} (E)$, and the
associated dipole rates, are largest for classes with large
particle-hole numbers. Once equilibration provides a sufficient
minimum value for the occupation probabilities of large-$m$ classes,
these classes are responsible for the bulk of dipole absorption. It
appears that at that time the excitation processes within the same
class described by $\rho^{\text{acc}}_{m m}(E)$ prevails over the
excitation generating additional particle-hole pairs. This situation
is reached at about $100$-MeV excitation energy in the upper left-hand
panel of Fig.~\ref{fig:reldiff}. The overall tendency of dominant
excitation of the large-$m$ classes would be amplified with every
dipole absorption process. However, each absorbed photon promotes the
nucleus to higher energy where the level densities $\rho_m(i, k)$ and,
thus, also the rates for equilibration become larger. For that same
reason, equilibration is faster for
\(\widetilde{\Gamma}_{\mathrm{dip}} = 20\)~MeV. Here the difference
between the underpopulation of small $m$ values and the overpopulation
of large $m$-values is less pronounced at $t = 5$ zs and has almost
disappeared at $t = 10$ zs. The center region of equilibrated
occupation probabilities where $R_m(k) \approx 0$ has the shape of a
stripe which runs almost parallel to but below the line $k(m)$ (not illustrated) defined
by the $k$ value where the density $\rho_m(0, k)$ versus $m$ has its
maximum. As time increases, that central equilibrated stripe becomes
wider, and it becomes more steep than the line $k(m)$.

Generally speaking, the occupation probabilities $P_m(0, k, t)$
deviate most strongly from equilibrium at short times ($t \leq 2$ zs)
and, with increasing time, tend towards equilibrium. That is expected
and is true for both values of the effective dipole
width. Equilibration becomes faster as energy increases while
saturation slows photon absorption. The difference between the
quasiadiabatic and the sudden regime is, therefore, manifest mainly at
short times and at comparatively low excitation energies and fades
away as the nucleus approaches saturation. We conclude that as long as
neutron emission is not taken into account, the sudden regime is quite
similar to the quasiadiabatic regime, except for the initial phase of
the process.

\subsubsection{Neutron evaporation}

\begin{figure}[htbp]
    \centering
    \includegraphics[width=\linewidth]{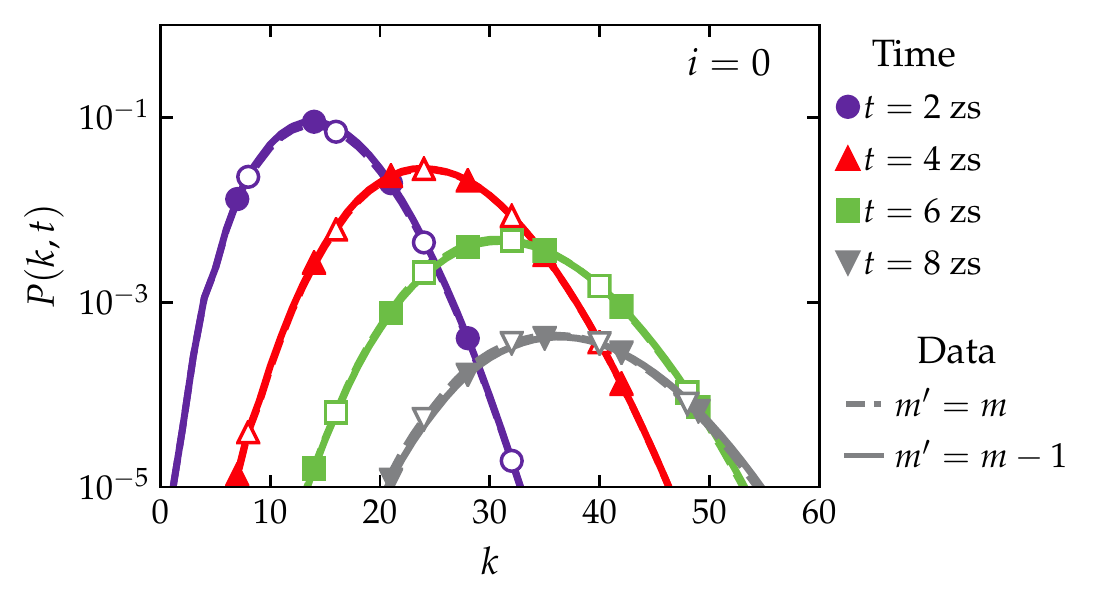}
    \caption[]{ Comparison of total occupation probabilities for target nucleus ($i=0$) summed over all particle-hole classes
      as a function of  the number $k$ of absorbed
      photons at times $t = $ 2, 4, 6, and 8 zs. We  calculate the neutron decay rates considering only $m'=m$ (dashed lines) or only $m'=m-1$ (solid lines). 
     We use
      \(\widetilde{\Gamma}_{\mathrm{dip}} = 5\)~MeV.}
    \label{neutron-comp}
\end{figure}

To include neutron evaporation we consider the target nucleus ($A =
100$, $i = 0$) plus three daughter nuclei with mass numbers $A=99$,
$A=98$, and $A=97$ ($i = 1, 2$, and $3$, respectively). We disregard
neutron emission of the last nucleus with mass number $A=97$ which,
thus, serves as a dump for the overall probability flow. 
  Our numerical results show that the contributions owing to $m'=m$ and
  to $m'=m - 1$ in Eq.~(\ref{gamma_n}) are almost equal.
  This is illustrated in Fig.~\ref{neutron-comp} presenting the
  respective total occupation probabilities as a function of the
  number of absorbed photons $k$ for $i=0$ and
  \(\widetilde{\Gamma}_{\mathrm{dip}} = 5\)~MeV. That comparison looks
  similar for $i=1,2,3$.

\begin{figure}[htbp]
    \centering
    \includegraphics[width=\linewidth]{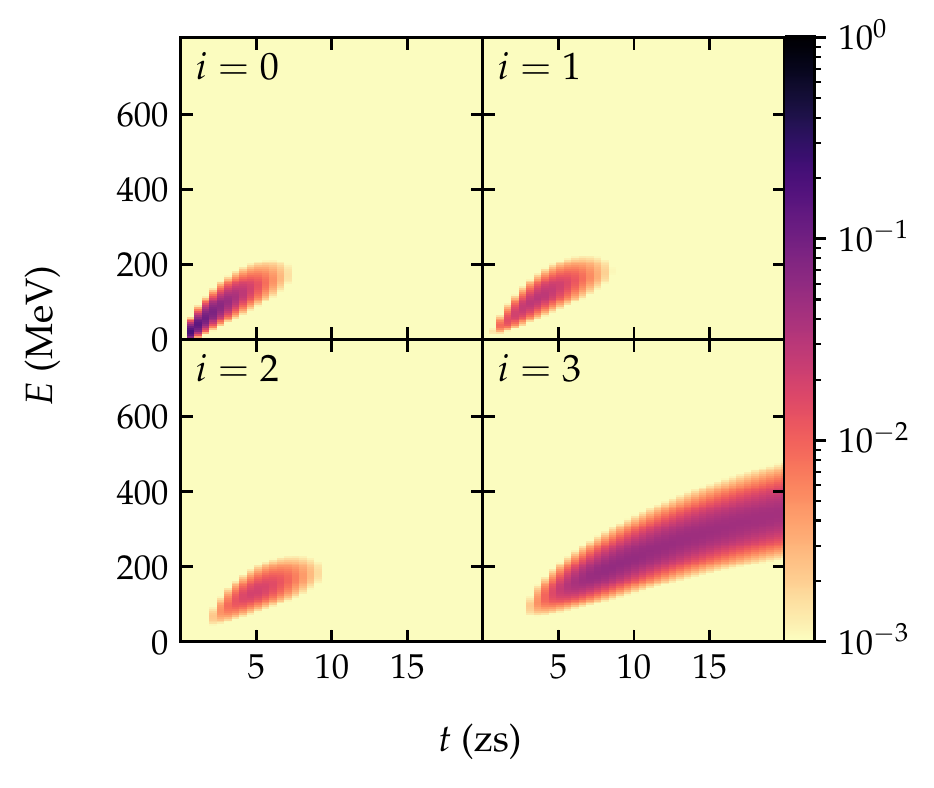}
    \caption[A=100, medium-weight, summed occupation probabilities
      with neutron evaporation]{Contour plots of the time-dependent
      occupation probabilities summed over all particle-hole classes
      as a function of excitation energy $E$ for target nucleus ($i =
      0$) and three generations of daughter nuclei ($i = 1, 2,
      3$). The parameters are the mass number \(A=100\), Eq.~(\ref{4})
      for the single-particle level density,
      $\widetilde{\Gamma}_{\mathrm{dip}} = 5 \)~MeV, $\tau =
      20$~zs and $\hbar \omega_0 = 5$~MeV. }
    \label{fig:4gens.summedprob.a100-5dip}
\end{figure}

\begin{figure}[htbp]
    \centering
    \includegraphics[width=\linewidth]{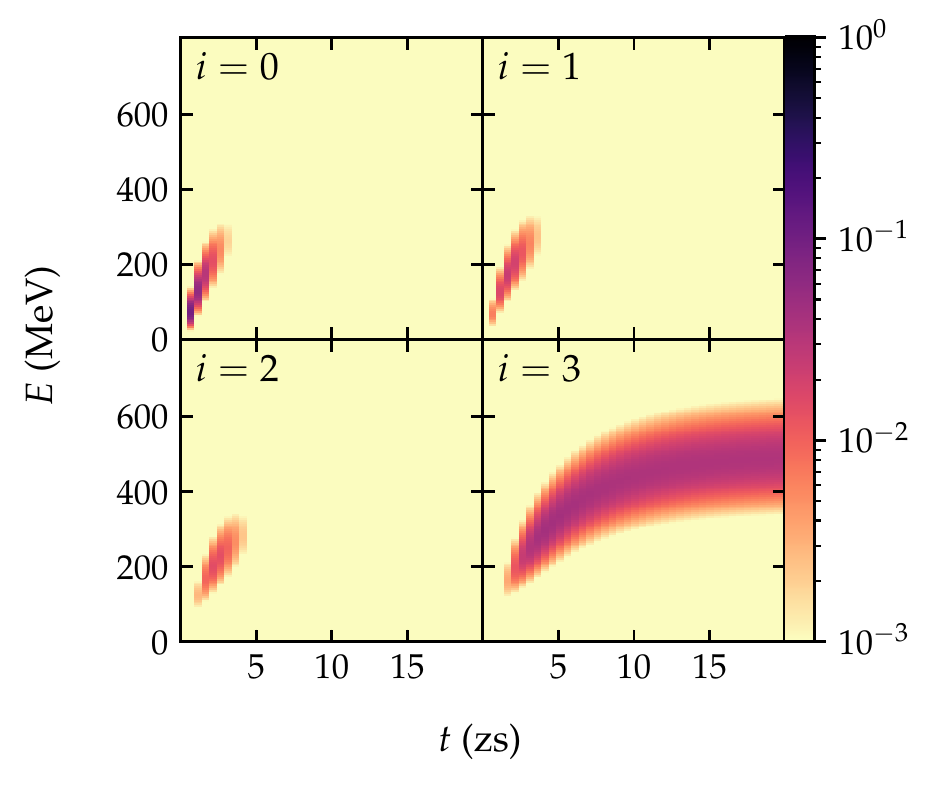}
    \caption[A=100, medium-weight, summed occupation probabilities
      with neutron evaporation]{The same as in
      Fig.~\ref{fig:4gens.summedprob.a100-5dip} for
      $\widetilde{\Gamma}_{\mathrm{dip}} = 20 \)~MeV.}
    \label{fig:4gens.summedprob.a100-20dip}
\end{figure}

In the
following we therefore simplify the calculation by considering the term $m'=m$
only.  The results for \(\widetilde{\Gamma}_{\mathrm{dip}} = 5\)~MeV
and \(\widetilde{\Gamma}_{\mathrm{dip}} = 20\)~MeV are presented in
Figs.~\ref{fig:4gens.summedprob.a100-5dip} and
\ref{fig:4gens.summedprob.a100-20dip}, respectively. Because the
relevant neutron evaporation decays take place within the first few
zs, we consider here a pulse duration time of $20$~zs. Both figures
are qualitatively similar to the quasiadiabatic case. Neutron
evaporation sets in at energies much lower than saturation,
interrupting the sequence of photoabsorption processes. In
Fig.~\ref{fig:4gens.summedprob.a100-5dip}, the occupation probability
of the target nucleus is lost by neutron decay already at \(t=7\) zs,
and the occupation probabilities of nuclei with $i = 1$ and $i = 2$
become significant. In comparison, for
\(\widetilde{\Gamma}_{\mathrm{dip}} = 20\)~MeV
(Fig.~\ref{fig:4gens.summedprob.a100-20dip}) the occupation
probability of the target nucleus reaches higher energies more quickly
and is completely depleted at $\approx$ 3.5 zs, which is about a
factor 2 sooner than for \(\widetilde{\Gamma}_{\mathrm{dip}} =
5\)~MeV. That trend is also seen for the nuclei with $i = 1$ and $i =
2$. By construction, in the present calculation neutron decay does not change particle-hole
class. The equilibration in the daughter nucleus is
therefore similar to the one  in the parent, modified only by the change of excitation
energy from the loss of one neutron. 

We expect qualitatively similar results for a longer chain of neutron
evaporation processes. Neutron decay prevents the nuclei in the chain
from reaching saturation. The length of the actual chain depends on
the duration of the laser pulse. In any case we expect that laser
irradiation leads to proton-rich medium-weight nuclei at high
excitation energy. The probability distribution of the nuclei in the
chain depends upon the parameters of the laser pulse. Once
experimental data become available, such details can be explored
further by calculations as performed in the present paper.

We compare our results for the chain of four nuclei with $i = 0, 1, 2,
3$ with corresponding results for the quasiadiabatic regime in
Ref.~\cite{Pal15}. These were done for
\(\widetilde{\Gamma}_{\mathrm{dip}} = 5\)~MeV but with a different
1p-1h density $\rho_1(0,1)$ that was taken from
Ref.~\cite{Obloinsk1986}. We adjust our calculations correspondingly.
For the target nucleus, the occupation probabilities are very similar
in the sudden and in the quasiadiabatic calculation (without
considering neutron evaporation). Neutron evaporation occurs slightly
faster in the quasiadiabatic regime. Inspection of the quantities
$\sum_m\Gamma_N(i,k,m)\rho_m(i,k)/\rho(i,k)$ and $\Gamma_N(i,k)$,
where $\rho(i,k)$ and $\Gamma_N(i,k)$ are the level density and total
neutron evaporation rate in Ref.~\cite{Pal15}, respectively, shows
that indeed between $k=10$ and $k=30$, neutron evaporation is stronger
in the quasiadiabatic regime. The difference is largest at $\approx$
$E=68$~MeV. 
In Ref.~\cite{Pal15}, tails of the level density up to an
excitation energy of 68 MeV were calculated using the Bethe formula
\cite{Bethe1936} while the central part was calculated using the
approach of Ref.~\cite{Pal13b} also employed here. Because of that
procedure the level density $\rho(i,k)$ of the quasiadiabatic
approach has a kink at $E=68$~MeV. Such a kink does not appear in the
present work where we extrapolate the level density in the tails. We
conclude that the observed difference in the neutron evaporation rate
is related to the different method used in calculating the level
densities at small energies. With this proviso we conclude that for
\(\widetilde{\Gamma}_{\mathrm{dip}} = 5\) MeV, the present calculation
confirms the previous results in Ref.~\cite{Pal15} for the
quasiadiabatic regime. It is clear, nevertheless, that the occupation
probabilities in the decay chain are sensitive to details of the
calculation such as the precise form of the level densities and the
manner in which photon absorption and stimulated photon emission
changes the occupation of the particle-hole classes.  Future experiments
are yet to confirm  the correct assumptions required for more quantitative estimates.

\section{Summary and Discussion \label{disc}}

Previous work on the laser-nucleus interaction~\cite{Pal14,Pal15} was
focused on the quasiadiabatic regime where the compound nucleus
equilibrates after each photoabsorption process. For the theoretical
modeling, it suffices to use the total level density at fixed
excitation energy. In the present paper we have investigated the
sudden regime where equilibration is incomplete. That regime requires
a more detailed modeling. We use classes of particle-hole states and
assume that within each class, equilibration is instantaneous. That
assumption is required to justify a statistical modeling and the use of
rate equations. The interaction between classes at the same excitation
energy leads to equilibration. Equilibration competes with multiple
photon absorption and induced photon emission. We also allow for
neutron evaporation feeding a chain of proton-rich nuclei.

In the absence of neutron evaporation and for a comparatively small
value \(\widetilde{\Gamma}_{\mathrm{dip}} = 5\) MeV of the effective
dipole width, equilibration competes successfully with dipole
absorption, and our results are in good agreement with those for the
quasiadiabatic regime of Ref.~\cite{Pal15}. For
\(\widetilde{\Gamma}_{\mathrm{dip}} = 20\) MeV, on the other hand, the
occupation probabilities of the particle-hole classes deviate markedly
from their equilibrium values in the beginning stages of the
multi-photon absorption process. In later stages, they approach the
equilibrium values, and the resulting excitation pattern becomes
qualitatively similar to that of the quasiadiabatic regime. That
happens before saturation (caused by the equality of the rates for
dipole absorption and induced dipole emission) limits the further
increase of excitation energy. Neutron evaporation actually sets in
long before saturation, depletes the target nucleus, and feeds a chain
of proton-rich nuclei. Repeated neutron evaporation somewhat decreases the excitation energy and slows down the path to saturation for each
nucleus in the decay chain.

 Throughout the paper we have neglected both fission and direct
emission of nucleons by photoabsorption into the continuum. As shown
in Ref.~\cite{Pal14}, these processes play only a minor role for
nuclei around $A = 100$, but may be competitive with neutron decay for
heavier nuclei. The effects of fission for $A=200$ were investigated in more
detail in Ref.~\cite{Pal15} for the quasiadiabatic regime. 
The effective charges of neutrons and protons
being nearly equal in magnitude, direct photoabsorption might still be
of interest also for medium-weigth nuclei, especially for \(\widetilde{\Gamma}_{\mathrm{dip}} = 20\)
MeV. That process would populate highly excited states not only in the chain
of proton-rich nuclei reached by neutron emission, but also in all
nuclei that lie between the valley of stability and nuclei in the
chain.

All our calculations were done for photons with energy $\hbar \omega_0
= 5$ MeV. Doubling that energy would lift it above the nucleon binding
energy. That would substantially increase direct photoabsorption
processes and might lead to a significant loss of mass.

The sudden regime is bounded by a regime where dipole excitation is so
strong that equilibration is altogether excluded. Then our assumption
that within every class of particle-hole states equilibration is
instantaneous fails. Multiple dipole excitation generates pairs of more or
less independent particle-hole states. It would be of substantial
interest to investigate the transition of the compound nucleus from a
strongly interacting system (realized in the adiabatic regime) to a
system of nearly independent particles (realized in the extreme sudden
regime of the laser-nucleus interaction).


\begin{acknowledgments}
This work is part of and supported by the DFG Collaborative Research
Center ``SFB 1225 (ISOQUANT)''. AP gratefully acknowledges support from the Heisenberg Program of the Deutsche Forschungsgemeinschaft (DFG).
\end{acknowledgments}

\bibliographystyle{apsrev}
\bibliography{disser}

\end{document}